\documentclass[10pt,twocolumn,twoside]{IEEEtran}

\usepackage{graphicx}
\usepackage{subfigure}
\usepackage{color}
\usepackage[utf8]{inputenc}
\usepackage{amsmath,amssymb,epsfig,theorem,url,cite,bm,soul}
\usepackage{multirow}
\usepackage{rotating}
\usepackage{wrapfig }

\usepackage{soul}
\usepackage{color}
\usepackage[usenames,dvipsnames]{xcolor}

\begin{document}

\title{Source Camera \textcolor{black}{Verification} for Strongly Stabilized Videos}

\author{Enes~Altinisik, H\"usrev~Taha~Sencar
}
\maketitle

\begin{abstract}

Image stabilization performed during imaging and/or post-processing poses one of the most significant challenges to photo-response non-uniformity based source camera attribution from videos.
When performed digitally, stabilization involves cropping, warping, and inpainting of video frames to eliminate unwanted camera motion.
Hence, successful attribution requires inversion of these transformations in a blind manner. 
To address this challenge, we introduce a source camera verification method for videos that takes into account spatially variant nature of stabilization transformations and assumes a larger degree of freedom in their search.
Our method identifies transformations at a sub-frame level, incorporates a number of constraints to validate their correctness, and offers computational flexibility in the search for the correct transformation.
The method also adopts a holistic approach in countering disruptive effects of other video generation steps, such as video coding and downsizing, for more reliable attribution. 
Tests performed on one public and two custom datasets show that the proposed method is able to verify the source of 23-30\% of all videos that underwent stronger stabilization, depending on computation load, without a significant impact on false attribution.

%
%

\end{abstract}

\section{Introduction}

Photo-response non-uniformity (PRNU) is an intrinsic characteristic of a digital imaging sensor 
that reveals itself as a unique and permanent pattern introduced to all media captured by the sensor.
The PRNU of a sensor is proven to be a viable identifier for source attribution,  
and it has been successfully utilized for identification and verification of the source of digital media.
In the past decade, various approaches have been proposed for reliable estimation, compact representation, and faster matching of PRNU patterns. 
These studies, however, mainly \textcolor{black}{feature} photographic images, and videos have been largely neglected.
In essence, steps involved in generation of a video are much more disruptive to PRNU pattern, and therefore its estimation from videos involves various additional challenges. 

PRNU is caused by variations in size and material properties among photosensitive elements that comprise a sensor. 
These essentially affect the response of each picture element under the same amount of illumination. 
Therefore the process of identification boils down to quantifying the sensitivity of each picture element.
This is realized through an estimation procedure using a set of pictures acquired by the sensor \cite{chen}.
To determine whether a media is captured by a given sensor, the estimated sensitivity profile, {\em i.e.,} the PRNU pattern, from the media in question is compared to a reference PRNU pattern obtained in advance using a correlation based measure, most typically using the peak-to-correlation energy (PCE) \cite{PCE}. 
The reliability and accuracy of the decision strongly depend on two main factors. 
First relates to the fact that the PRNU bearing raw signal at the sensor output has to pass through several steps of in-camera processing before a media is generated, which may further be \textcolor{black}{subject} to some out-of-camera processing. 
These processing steps will have a weakening effect on the inherent PRNU pattern.
Second, and more critically, it relies on preserving the element-wise correspondences between the reference pattern and the PRNU pattern estimated from the media in question.

\begin{figure}
	\centering
	\includegraphics[width=1\columnwidth]{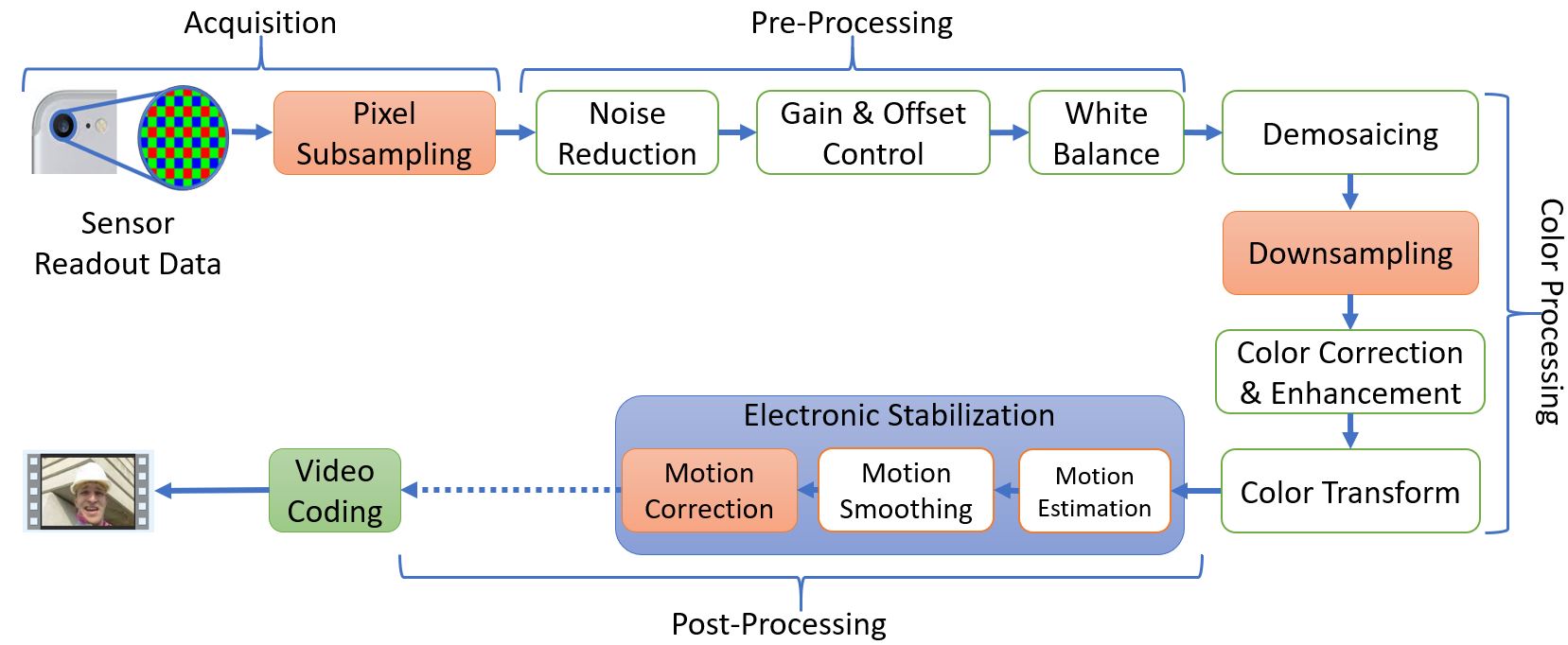}
	\caption{The processing steps involved in the \textcolor{black}{imaging pipeline of a camera} when generating a video. The highlighted boxes are specific to video capture.}
	\label{fig:pipeline}
\end{figure}

The imaging sub-systems used by digital cameras largely remain proprietary to the manufacturer; therefore, it is quite difficult to access their details.
At a higher level, however, the imaging pipeline in a camera includes various stages, such as acquisition, pre-processing, color-processing, post-processing, and image and video coding.
Figure \ref{fig:pipeline} shows basic processing steps involved in capturing a video, most of which are also utilized during the acquisition of a photograph \cite{corcoran2016consumer}.
When generating a video, an indispensable processing step is the downsizing of the full-frame sensor output to reduce the amount of data that needs processing.  
This may be realized at acquisition by subsampling pixel readout data (trough binning \cite{zhang2018pixel} or skipping \cite{guo2018efficient} pixels), as well as during color processing by downsampling color interpolated image data.
Another key processing step employed by modern day cameras at post-processing stage is the electronic image stabilization which aims at compensating camera shake related blur.
It must be noted that even such post-processing may include operations, such as cropping and scaling, that result with further downsizing of pictures.
At last, the sequence of post-processed pictures are encoded into a standard video format for effective storage and transfer.


Crucially, video generation involves three additional steps as compared to the generation of photos, including downsizing, stabilization, and video coding. 
When combined together, these operations have a significant adverse impact on PRNU estimation in two main respects.
These relate to geometric transformations applied during downsizing and stabilization and the information loss largely caused by resolution reduction and compression.
\textcolor{black}{These detrimental effects could be further compounded by out-of-camera processing.  
Video editing tools today provide a wide variety of sophisticated processing options over cameras as they are less bounded by computational resources and real-time constraints.
}
Therefore, estimation of a sensor's PRNU from videos requires addressing these challenges.

A number of approaches have already been proposed to address these problems with different degrees of effectiveness.
Earlier works mostly focused on coping with compression as it is typically more lossy for videos than images.
To obtain a better PRNU estimate, Chen {\em et al.} \cite{PRNUvideoJessica} considered removal of periodic structures in the PRNU pattern that arise due to block-based operation of encoder; Hyun {\em et al.} \cite{MACEfilter} introduced the use of minimum average correlation energy filter to better suppress compression related noise during matching; and Chuang {\em et al.} \cite{IPBdiff}, observing intracoded (I) frames yield more reliable PRNU patterns than predicted (P and B) frames, suggested weighting I frames more heavily during estimation.
More recently introduced approaches proactively intervene in the decoding process to compensate for the deblocking filter \cite{eusipco18} and incorporate macroblock level compression information into estimation process \cite{H246H265}.

The weakening of effect of downscaling on PRNU pattern was long observed. 
In \cite{ei2019}, it is shown that downsizing high-resolution sensor output by a factor higher than six removes almost all traces of PRNU pattern in a video, even for very high quality videos, when the specifics of downsizing method are not known.
Moreover, downsizing is also a concern because of the geometric distortions it introduces on the PRNU pattern as cameras can capture media at a variety of resolutions.
Therefore, mismatches in resolution between a reference PRNU pattern and a media need to be taken into consideration. 
References \cite{ei2019}, \cite{piva}, and \cite{taspinar2019source} examined in-camera downsizing behavior to enable reliable matching of PRNU patterns at different resolutions.  

In regards to preserving synchronicity between PRNU patterns, the most significant challenge is posed by image stabilization.
When performed electronically, stabilization requires estimating and removing undesired camera motion due to handheld shooting or other vibrations.
This involves the application of geometric transformations to align successive frames in a video with respect to each other.
From the PRNU estimation standpoint, this requires registration of PRNU patterns by inverting transformations applied to each frame in a blind manner.
Ultimately, the difficulty of this task depends on the type of stabilization transformations applied to video frames. 
\textcolor{black}{For the sake of clarity, in this work, we refer to videos that can be attributed through application of frame-level affine transformations as {\em weakly stabilized} videos and those that require more complicated transformation inversion settings as {\em strongly stabilized} videos.}


The approaches proposed so far to deal with stabilization focused on a variety of aspects including determining the presence of stabilization in a video \cite{taspinar2016} as well as verifying the source of a video by evaluating frame-level matches \cite{piva} and obtaining a reference PRNU from weakly stabilized videos \cite{luisapaper} under an affine transformation model. 
\textcolor{black}{Our main contribution in this work lies in extending source verification capability, which assumes the presence of camera's reference PRNU pattern, to strongly stabilized videos}.
Essentially, inspired by the approach introduced in \cite{ref-Content2009}, many proposed stabilization methods effectively involve application of spatially varying warps during stabilization. 
By taking this into account, our work departs from earlier attribution approaches in its premise that stabilization transformations may exhibit locality and not necessarily be applied at the frame level as assumed by prior work.

More specifically, our proposed method for verifying source of stabilized videos differs from earlier methods in two main aspects.  
First, in countering the variant nature of stabilization transformations, our method operates on blocks of frames, rather than on individual frames.
To avoid blocks whose content partially underwent multiple warpings,
we evaluate the coherence of matching results obtained at the block and sub-block levels.
Second, in reverting the transformation applied to each block, we consider projective transformations, as opposed to affine transformations, which provide wider flexibility in identifying the unwanted motion removed by the stabilization.

\textcolor{black}{
In summary, we make the following contributions:
\begin{itemize}
    \item 
    We provide an in-depth examination of the challenges posed by modern stabilization approaches on video source camera attribution. 
    Our method \textcolor{black}{proposes} to address these challenges by taking into account the spatially variant nature of stabilization transformations and allows a larger degree of freedom in identifying the applied transformation by assuming a projective motion model rather than the commonly adopted affine model.
%
    \item 
    We present algorithms to alleviate the computational burden of searching stabilization transformation parameters. 
    \textcolor{black}{To this purpose}, we \textcolor{black}{propose} different approaches based on hierarchical grid partitioning of the transformation space and \textcolor{black}{analyze} their complexity. 
    \item We introduce measures to validate identified transformations as a necessary step to eliminate spurious matches arising due to incorrect transformations and demonstrate their effectiveness.
    \item 
    We investigate the effect of performing transform inversion in spatial and PRNU pattern domains as the latter offers a significant computational gain
    over the former during transform inversion.  
    We also explore the smallest possible PRNU block size that could be used for attribution of video frames. 
    \item Our approach is holistic as it utilizes findings of earlier works to differentiate between non-stabilized \cite{taspinar2016}, weakly stabilized \cite{piva, luisapaper}, and strongly stabilized videos and incorporates methods on mitigation of video compression effects \cite{H246H265} and downsizing behavior \cite{ei2019} into its operation.
    \item The proposed method is validated on three datasets. These include the publicly available VISION dataset \cite{dataset} and two newly created custom datasets. One of the newly generated datasets includes media captured by two iPhone camera models (the iPhone SE-XR dataset) and the other includes a set of externally stabilized videos using the Adobe Premiere Pro video processing tool (the APS dataset)\footnote{\textcolor{black}{The two newly generated datasets and the implementation for the proposed approach can be obtained at https://github.com/VideoPRNUExtractor following final modifications, prior to publication of this manuscript.}}. 
    \item We show that the proposed method is able to correctly attribute 23-30\% of all strongly stabilized videos in the three datasets, depending on the size of the transform search space, without any false attributions while utilizing only 10 frames from each video.
    %
\end{itemize}
}


In the next section, we describe how image stabilization is performed and provide an overview of proposed approaches for source attribution on stabilized videos. 
In section \ref{sec:Chlnge}, we address challenges in attributing stabilized videos with excessive camera motion.
Details of our method are described in Section \ref{sec:Mthd} and performance results are presented in Section \ref{sec:experiments}.
Finally, our discussion on the results is given in Section \ref{sec:Conc}.

\section{Video Stabilization} \label{sec:Stab}


With the increasing processing power built into cameras and the advances in lens technologies, 
increasingly more powerful stabilization solutions have become available on cameras. 
In essence, there are two primary approaches to image stabilization.
The first one is the optical stabilization.
In this approach, stabilization is performed mechanically through the use of hardware based mechanisms, and the movement of the camera is not fully transferred to the video.
Rather, it is absorbed by moving the lens or the imaging sensor to counter the unwanted motion.
Since optical stabilization preserves pixel-to-pixel correspondences in successive frames, 
it does not obstruct PRNU based source attribution. 

The other approach is the digital stabilization where frames captured by the sensor are moved and warped to align with one another through processing.
With this approach, the movement of the camera is estimated either from sequence of frames or using available sensors on the device.
Then, corrective stabilization transforms associated with the estimated motion are determined, and each frame is transformed and saved accordingly. 
\textcolor{black}{
This form of stabilization can also be applied externally by aligning frames extracted from a video file with greater freedom in processing.
In both cases, however, this frame level processing introduces asynchronicity among PRNU patterns of consecutive frames in a video which can be detrimental to PRNU based source attribution.}

Attribution of digitally stabilized videos thus requires understanding the specifics of how stabilization is performed. 
This, however, is a challenging task as inner workings and technical details of processing steps of camera pipelines are usually not revealed.
In fact, both stabilization approaches can be deployed together in a camera for more effective results \cite{apple}.
Further, even in the absence of abrupt camera motion the vibrations caused by physiological hand tremor may induce undesirable blur in videos \cite{elTitremesi}; therefore, when performed digitally, stabilization effects can potentially be present in most videos.
At a high level, the three main steps of digital stabilization involve camera motion estimation, motion smoothing, and 
alignment of video frames according to the corrected camera motion.   
Motion estimation is performed either by describing the geometric relation between consecutive frames through a parametric model
or through tracking key feature points across frames to obtain feature trajectories \cite{Xu-FastFeature}, \cite{grundmann-AutoDirected}.
With sensor-rich devices such as smartphones and tablets becoming the primary camera, data from motion sensors are also utilized to improve the estimation accuracy \cite{patent_stab_gyro}.
This is followed by application of a smoothing operation to estimated camera motion or obtained feature trajectories to eliminate the unwanted motion. 
Finally, each frame is warped according to the smoothed motion parameters to generate the stabilized video.  
The most critical factor in stabilization depends on whether the camera motion is represented by a two dimensional (2D) or three dimensional (3D) model. 
Early methods mainly relied on the 2D motion model that involves application of full-frame 2D transformations, such as affine or projective models, to each frame during stabilization.
Although this motion model is effective in scenes far away from camera where parallax is not a concern, it does not generalize to more complicated scenes captured under spatially variant camera motion. 
To overcome 2D modelling limitations, more sophisticated methods considered 3D motion models. 
However, due to difficulties in 3D reconstruction, which requires depth information, these methods introduce simplifications to 3D structure and rely heavily on the accuracy of feature tracking \cite{ref-Content2009, LIU-SteadyFlow, 360Kopf, Wang-HiQuRealTime}.
Most critically, these methods involve the application of spatially-variant warping to video frames in a way that preserves the content from distortions introduced by such local transformations. 
This poses a significant complication to PRNU based source attribution, as for each frame it requires determining the inverse warping parameters at a local level, and not globally. 

\subsection{Work on Attribution of Stabilized Videos}

In essence, digital image stabilization tries to align content in successive frames through geometric registration.
Depending on the complexity of camera motion during capture, this may include application of a simple Euclidean transformation (scale, rotation, and shift applied individually or in combination) to spatially-varying warping transformation in order to compensate for any type of perspective distortion.
Because these transformations are applied on a per-frame basis and the variance of camera motion is high enough to easily remove
pixel to pixel correspondences among frames, alignment or averaging of frame level PRNU patterns will not be very effective in estimating a reference PRNU pattern. 
Therefore, performing source attribution in stabilized video requires determining and inverting those transformations applied at the frame level. 


Source attribution under geometric transformations was studied earlier to verify the source of transformed images when the reference PRNU pattern is available.
Considering scaled and cropped photographic images, Goljan {\em et al.} \cite{JessicaCrop} proposed a brute force search for the geometric transform parameters. 
For this, the PRNU pattern obtained from the image in question is upsampled in discrete steps and matched with the reference PRNU at all shifts. 
The parameters that yield the highest PCE are identified as the correct scaling factor and the cropping position. 
More relevantly, by focusing on panoramic images, Karakucuk {\em et al.} \cite{karakucuk2015} investigated source attribution under more complex geometric transformations.
Their work showed the feasibility of estimating inverse transform parameters considering projective transformations.  


In the case of stabilized videos, Taspinar {\em et al.} \cite{taspinar2016}
proposed determining the presence of stabilization in a video by extracting reference PRNU patterns from the beginning and end of a video and by testing the match of the two patterns. 
If stabilization is detected, one of the I frames is designated as a reference and other I frames are aligned with respect to it through a search of inverse affine transforms to correct for the applied shift and rotation.
The pattern obtained from the aligned I frames is then matched with a reference PRNU pattern obtained from a non-stabilized video by performing another search. 
The approach is validated on manually stabilized videos using FFMPEG deshaker.


Iuliani {\em et al.} \cite{piva} introduced another source verification method similar to \cite{taspinar2016} by additionally assuming the reference PRNU pattern might have been obtained from photos as well as from a non-stabilized video. 
That is, the video in question may have a different resolution than the reference PRNU pattern, and this mismatch in scales need to be taken into account during matching.
To perform verification, 5-10 I frames are extracted and corresponding PRNU patterns are aligned with the reference PRNU pattern by searching for the correct amount of scale, shift and cropping applied to each frame. 
Those frames that yield a matching statistic above some predetermined PCE value are combined together to create an aligned PRNU pattern. 
\textcolor{black}{Tests performed on videos captured by 8 cameras in the VISION dataset using first 5 frames of each video revealed that 86\% of videos captured by cameras that support stabilization in the reduced dataset can be correctly attributed to their source with no false positives.}
They showed that the method is also effective on a subset of videos downloaded from YouTube with an overall accuracy of 87.3\%.

In \cite{luisapaper}, Mandelli {\em et al.} introduced a method for estimating the PRNU pattern considering weakly stabilized videos. 
In this approach, a reference for alignment is generated from a set of frames.
For this, PRNU estimates obtained from each frame is matched with other frames in a pair-wise manner to identify those translated with respect to each other. 
Then the largest group of frames that yield a sufficient match are combined together to obtain an interim reference PRNU pattern and remaining frames are aligned with respect to this pattern. 
Alternatively, if the the reference PRNU pattern at a different resolution is already known, then this is used as a reference and PRNU patterns of all other frames are matched by searching for transform parameters using particle swarm optimization.
They observed that for weakly stabilized videos, rotation can be ignored to speed up the search. 

When performing source verification, sensor's PRNU pattern is first estimated from a weakly stabilized flat and still content videos as described above. 
For verification, five I frames extracted from the stabilized video are matched to this reference PRNU 
pattern considering a scaling by a factor of 0.99 to 1.01, rotations of -0.15 to 0.15 radians, and all possible shift positions. 
If the resulting PCE values for at least one of the frames is observed to be higher than a threshold, a match is assumed to be achieved.
\textcolor{black}{Results obtained on the VISION dataset show that the method is effective in successfully attributing 71\% and 77\% of videos captured by cameras that support stabilization with 1\% false positive rate, respectively, when 5 and 10 I frames are used while excluding the first I frame as it is less likely to be stabilized.
Alternatively, if the reference PRNU pattern is extracted from photos, rather than flat videos, under the same conditions attribution rates increase to 87\% for 5 frames and to 91\% for 10 frames.}

We next describe additional challenges involved in dealing with stabilized videos captured under more severe camera motion and introduce our approach that is complementary to above methods. 


\section{Additional Challenges}\label{sec:Chlnge}
The difficulty of inverting per-frame warping transformations is further exacerbated by additional factors.
Video frames have lower resolutions than the full-sensor resolution typically used for acquiring photos.
Therefore, a reference PRNU pattern estimated from photos provides a more comprehensive characteristic, 
but its use for video source verification potentially introduces a mismatch with the size of video frames.
Essentially, downsizing operation in a camera involves various proprietary hardware and software mechanisms that crucially involve sensor cropping and resizing.
Performing source attribution on stabilized video requires determining such device dependent parameters in advance. 
When this is not possible, the search for inverse warping transformations has to incorporate the search for these parameters as well.

Lower PCE values observed in matching PRNU patterns obtained from videos, as compared to those from photos, yields another complication. 
This decrease in PCE values is primarily caused by downsizing operation and video compression. 
When the PRNU pattern is estimated from multiple video frames, downsizing can be ignored as a factor as long as the resizing factor is higher than $\frac{1}{6}$ and the compression becomes the main concern \cite{ei2019}. 
As demonstrated in \cite{H246H265}, at medium to low compression levels, average PCE values drop significantly as compression gets more severe. 
Accordingly, reference patterns extracted from 36 raw videos captured by 28 cameras that are downsized in camera by a factor of four and compressed at 2 Mbps, 900 Kbps and 600 Kbps bit rates, respectively, yielded average PCE values of 2000, 300, and 40.
Alternatively, when PRNU patterns from video frames are individually matched with the reference pattern ({\em i.e.}, frame-to-reference matching), even downsizing by a factor of 2 causes significant reduction in measured PCE values \cite{down2}.
Tests performed on 14 videos captured by 7 cameras at a resolution of $1920 \times 1080$ pixels by performing frame-to-reference matching revealed that resulting PCE values are mostly around 20, and below 40 for almost all frames. 
This introduces a significant challenge in the search of the correct transformation parameters.


Another issue concerns the difficulty of setting a decision threshold for matching. 
Large scale tests performed on photographic images show that setting the PCE value to 60 as a threshold yields extremely low false-matches when the correct-match rate is quite high. 
In contrast, as demonstrated in the results of earlier works, where decision thresholds of 40-100 \cite{piva} and 60 \cite{luisapaper} are utilized when performing frame-to-reference matching, such threshold values on video frames yield much lower attribution rates.

Some of the in-camera processing steps introduce artefacts that obstruct correct attribution. 
The biases introduced to PRNU estimate by the demosaicing operation and blockiness caused by compression are known to introduce periodic
structures onto the estimated PRNU pattern. 
These artefacts can essentially be treated as pilot signals to derive clues about the transformation history of media after the acquisition. In fact for the case of photos, the linear pattern associated with the demosaicing operation has shown to be effective in determining the amount of shift, rotation, and translation, with weaker presence in newer cameras \cite{Goljan-LP}. 
In the case of videos, the linear-pattern is observed to be even weaker most likely due to application of in-camera downsizing and 
more aggressive compression of video frames as compared to photos.
Therefore, it cannot be reliably utilized in identifying global or local transformations.
In a similar manner, since video coding uses variable block sizes determined adaptively during encoding, blockiness artefact is also not useful in reducing the computational complexity of determining the warping transformation. 


Finally, the first frame of a video can be thought to be less affected from stabilization as most motion smoothing methods correct motion with respect to a reference frame \cite{ilkIFrameNonStabil}, which is typically the first frame.
\textcolor{black}{In \cite{luisapaper}, it is reported that the first frame of videos in the VISION dataset are mostly non-stabilized.
Our measurements also verify that, assuming a translation motion model, the first frame in 209 of those 257 videos in the VISION dataset yields a PCE value higher than 60. 
However, this finding does not hold for any of the videos in the newly generated iPhone SE-XR dataset. 
Hence, for some cameras, it seems stabilization gets activated when the camera is set to video mode, even before recording starts; therefore, the first frame cannot be solely relied on as the basis of attribution.
}






\section{Proposed Method} \label{sec:Mthd}

\begin{figure}[!h]
	\centering
	\includegraphics[width=1\columnwidth]{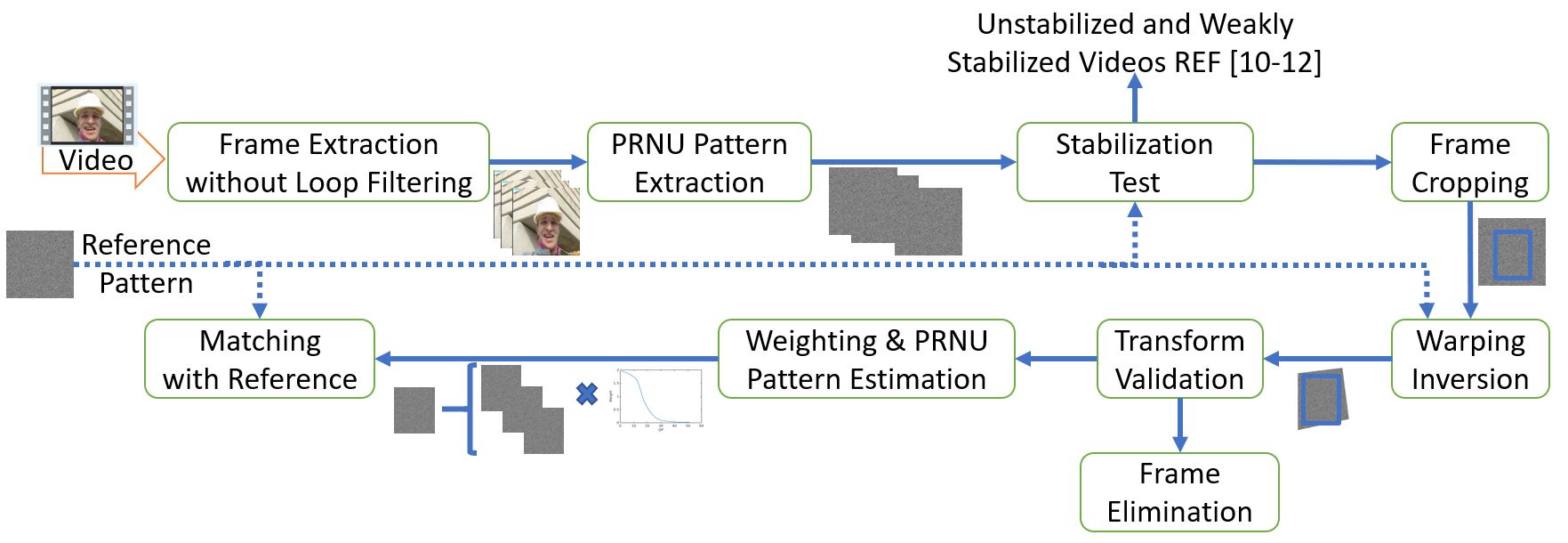}
	\caption{Source camera verification steps for stabilized videos.}
	\label{fig:process}
\end{figure}

Our approach to attribution of stabilized videos assumes a source verification setting where a given video is matched against a known camera. That is, the reference PRNU pattern is assumed to be available.
Our method comprises seven main steps.
First, the bitstream is decoded into video frames while compensating for the effects of a filtering procedure applied at the decoder ({\em i.e.,} the loop filter) to reduce coding artefacts.
Then, a PRNU pattern is extracted from each extracted frame.
Before the analysis, the video is also tested for the severity of stabilization to eliminate unstabilized and weakly stabilized videos which can be attributed by existing methods. 
This is followed by cropping out smaller blocks from each PRNU pattern to cope with spatially variant nature of stabilization transformations. 
A search is performed to identify transformation parameters for each PRNU block along with a validation step to prevent incorrect inversions.
The inverse-transformed blocks are then combined together by a weighting procedure that takes into account the compression level of each block.  
The estimated PRNU pattern is finally compared against the reference PRNU pattern to evaluate the match.
Figure \ref{fig:process} presents the sequence of attribution steps. 

\subsection{Loop filter compensation}

Compression is the last step in the video generation pipeline; therefore, video coding related artifacts must first be mitigated to reliably revert stabilization transformations. 
Among such artifacts the most detrimental is caused by filtering of compressed video frames.
Essentially, block-wise quantization of frame data during encoding introduces a blocking effect across block boundaries. 
To suppress these coding related visual artefact, H.264 and H.265 codecs incorporate filtering procedures both at the encoder and decoder.
While this improves visual quality of resulting video significantly, it also weakens the inherent PRNU pattern.
This weakening gets further emphasized at increasing compression levels. 
To address the disruptive effects of this filtering operation, \cite{H246H265} introduced a compensation method by modifying the decoder's operation. 
Results on test videos revealed that this method yields an improvement in measured PCE values with a three times average increase.
Hence, we utilize this method to compensate for the effects of the filtering process when extracting video frames from the bitstream. 

\subsection{Frame-wise PRNU Extraction}
Following the extraction of video frames, the process for inverting stabilization transformations starts. 
Since transformations are performed in the spatial domain, the search for the unknown transformation for each frame has to be ideally performed in the spatial domain where the correct transformation is validated based on the match of the estimated PRNU with the reference PRNU.  
That is, inverse transformation in spatial domain has to be followed by PRNU estimation. 
This order of operations, however, involves a significant amount of computation because transformation parameters are determined through a brute-force search and the search space for the parameters can be quite large.
Due to this complexity, earlier work \cite{taspinar2016,luisapaper,piva} changed the order of operations and searched for inverse transformation in the PRNU domain, rather than in the spatial domain, which is performed much faster as PRNU estimation is performed only once.
However, since the PRNU estimation operation is not of linear nature, this change in order is likely to introduce degradation in performance. 
Further, it must be noted that a geometric transformation also involves an interpolation operation as transformed coordinates will not correspond to grid positions in the original frame and missing values at those grid locations must be interpolated. 
Such interpolation will act as another disturbance on the underlying PRNU pattern. 

To determine the overall impact of performing a search in PRNU domain on performance, we performed a test.
\textcolor{black}{To this purpose}, we utilized videos taken under controlled conditions, by turning off stabilization and under low compression setting, using a custom camera application for Android mobile operating system \cite{ei2019}.
The test included 1000 frames from videos taken by 6 cameras and the corresponding reference PRNU patterns.
For each frame, we first evaluated the match with the reference PRNU pattern in terms of the PCE metric and determined that the average value for all frames is $219$.
To measure the impact of transformation related interpolation, we applied a random transformation and its inverse consecutively to each frame and
computed the match of estimated PRNUs with reference patterns.
Our evaluation of various widely used resampling methods, including the nearest neighbor, bicubic and bilinear interpolations, revealed that the nearest neighbor method induces the least distortion on the estimated PRNU pattern with the overall average dropping to $167$.
Finally, we applied the same sequence of random transformations to each frame, estimated PRNU patterns, inverted the transformation and re-evaluated the match with the reference PRNU which yielded the average of $162$.
The resulting PCE values show that search of parameters in the PRNU domain will potentially yield acceptable results in most cases.

\textcolor{black}{To exploit this computational advantage, in our method, we also perform a search for inverse transform parameters in the PRNU domain. 
Hence, full-frame PRNU patterns are extracted individually from loop-filter compensated video frames using the basic method introduced in \cite{chen}.}


\subsection{Stabilization Testing}

The steps involved in the attribution of a stabilized video are computationally intensive.
To effectively deal with this complexity, the level of stabilization applied to a video and how it is performed must also be taken into account. 
Therefore, rather than assuming that a video has undergone severe stabilization, it must first be checked for traces of weak stabilization
by assuming an affine model for camera motion.
Those videos can be attributed using earlier proposed approaches \cite{piva, luisapaper}, and only the remaining videos must be kept for further analysis considering more complex stabilization settings. 

In line with this thinking, we perform two tests to eliminate unstabilized and weakly stabilized videos from further testing.
To achieve this goal, we first apply a test, $stb_{chk}$, to verify the presence of a stabilization in a video. 
\textcolor{black}{This is realized by estimating two reference patterns from the first and last third parts of a video by utilizing the basic method \cite{chen} on loop-filter compensated video frames and evaluating their match.}
Videos determined to be stabilized are then subjected to another test, $stb_{lite}$, to identify weakly stabilized videos. 
This test is performed by geometrically aligning PRNU patterns of 10 I frames following the first I frame  with respect to the reference pattern through a search of affine transformation parameters.
Those frames that yield a PCE value of 38 after transform inversion are combined together to obtain a PRNU estimate as performed by \cite{piva}. 
If the resulting estimate yields a sufficient match, the test is considered a positive confirmation of weak stabilization. 
Videos that yield low values on both tests are kept for further analysis.


\subsection{Frame Cropping}
The most prominent stabilization approach involves application of spatially varying transformations to each frame rather than 
applying a global transformation.  
In its most simplest form, this reduces to splitting a frame into a grid and stabilizing each grid block locally 
where block sizes can be as small as $64 \times 36$ pixels \cite{ref-Content2009} or $40 \times 40$ pixels \cite{Wang-HiQuRealTime}.
With any stabilization approach, however, it is safe to assume that there will be some locality that has undergone a specific geometric transformation. 

\begin{figure}[htbp]
  \begin{minipage}[b]{0.48\columnwidth}
    \centering
    \includegraphics[width=1\columnwidth]{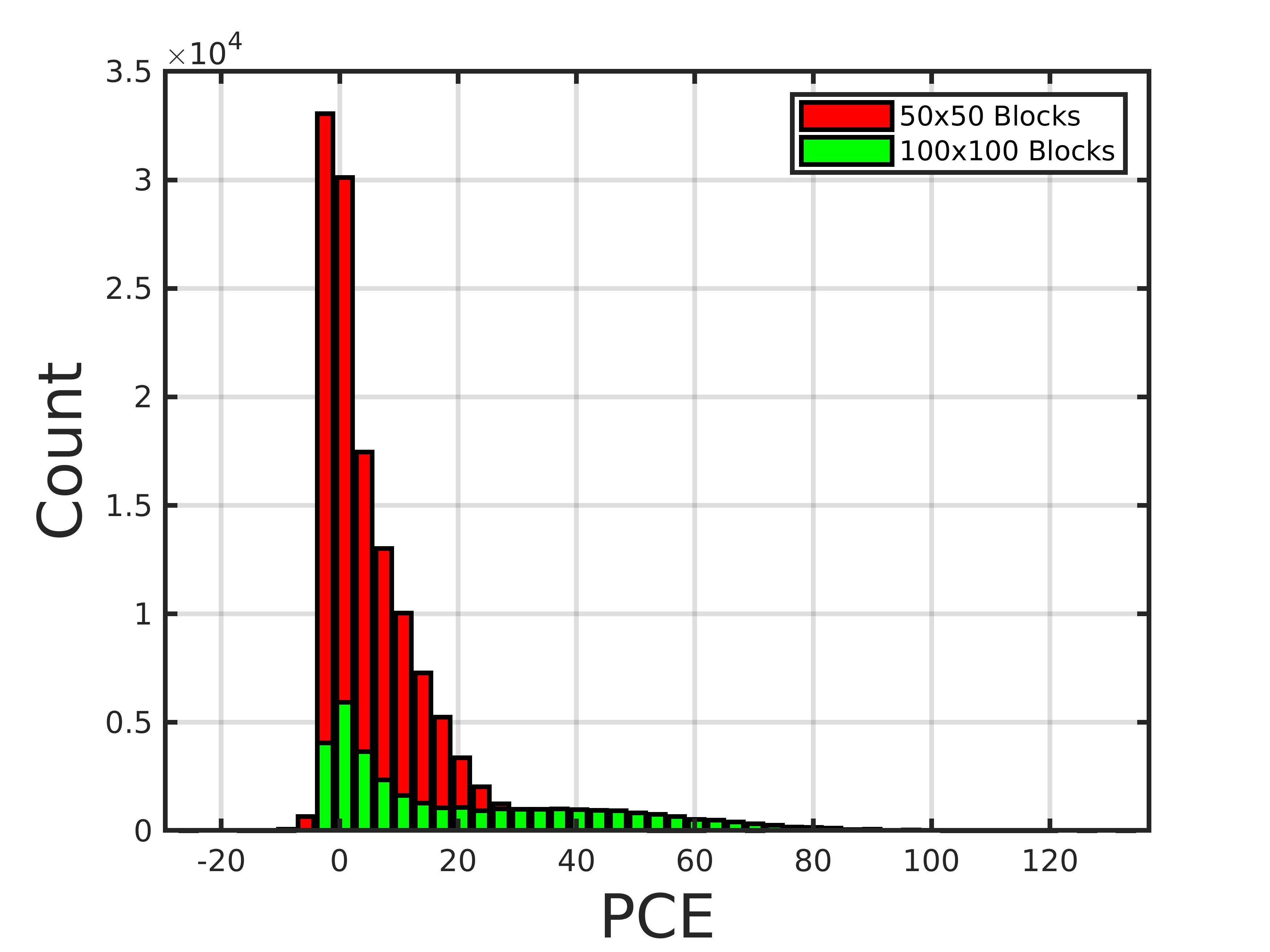}
    \centerline{(a)}
  \end{minipage}
  \begin{minipage}[b]{0.48\columnwidth}
    \centering
    \includegraphics[width=1\columnwidth]{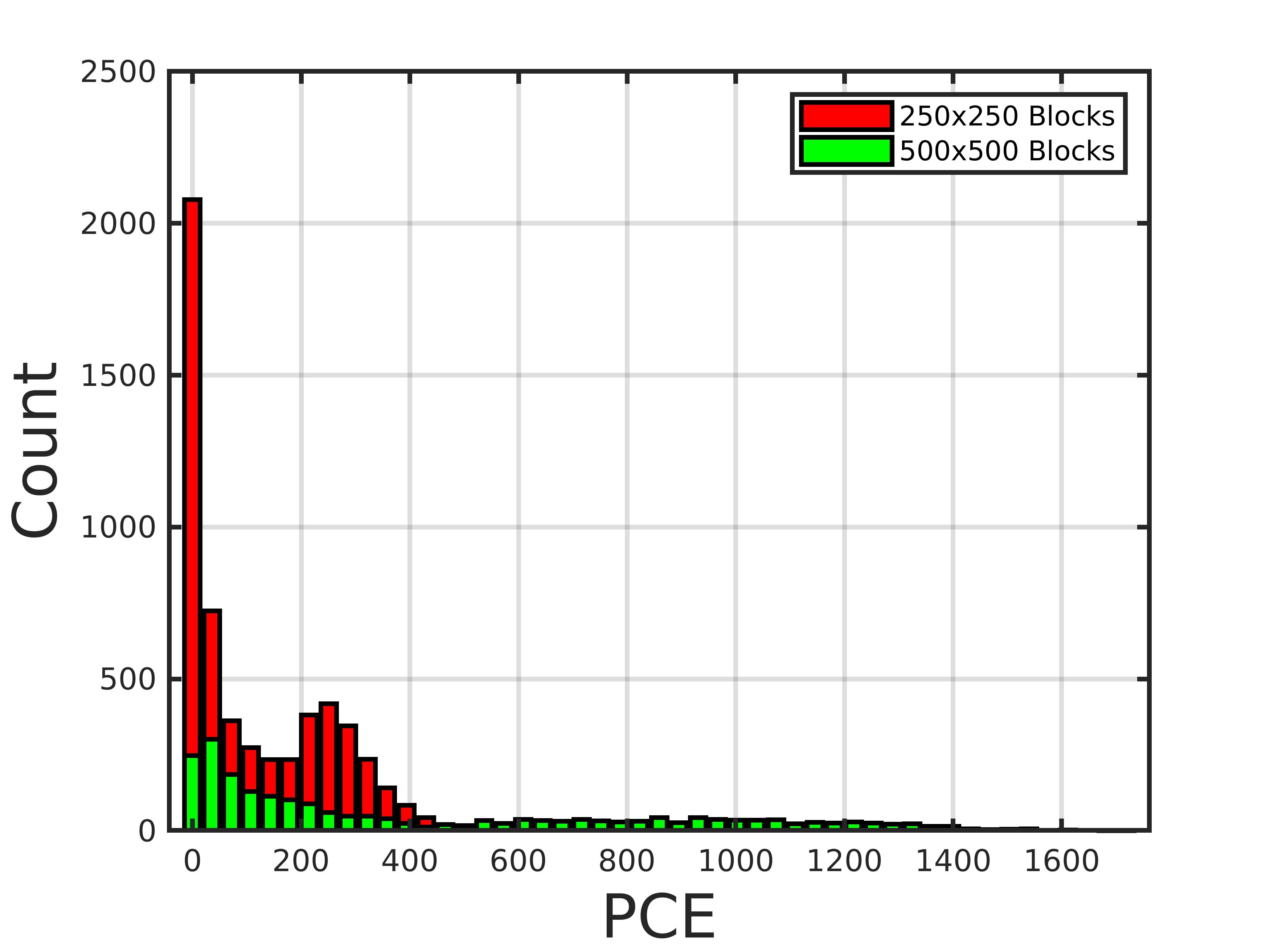}
    \centerline{(b)}
  \end{minipage}
  \caption{Histogram of PCE measurements obtained by matching (a) $50 \times 50$ and $100 \times 100$ (b) $250 \times 250$ and $500 \times 500$ sized blocks to corresponding blocks in the reference PRNU pattern.}
  \label{fig:boyut}
\end{figure}

Therefore, the size of blocks that needs to be used during search for inverse transformation parameter must be determined.
We performed tests to determine the smallest block size in a video frame that will yield meaningful PCE measurements.
\textcolor{black}{To this purpose}, we used seven unstabilized videos taken by different cameras with known reference PRNUs. 
The videos were compressed at the lowest possible compression and were captured indoors while the camera is moving \cite{ei2019}. 
In each frame, we cropped blocks of varying size, estimated the PRNU, and evaluated the match with the corresponding block in the reference PRNU.
Figure \ref{fig:boyut} provides histograms of measured PCE values when block size is set to $50 \times 50$, $100 \times 100$, $250 \times 250$ ve $500\times 500$ pixels.
As it can be seen from these results PRNU blocks with sizes of $50 \times 50$ and $100 \times 100$  do not yield reliable measurements where most PCE values are much lower than the commonly accepted threshold value of $60$. 
Even at the block size of $250 \times 250$ a significant number of blocks do not yield sufficiently high PCE values.
Therefore, in our method, we utilize a block size of $500\times500$ but also incorporate results of $250\times250$ sub-blocks when identifying transformation parameters. 
%

The other issue concerns the selection of the block location in each video frame.  
In our method, we select the $500\times500$ blocks at the center of each frame. 
This is primarily because of three reasons.
First, the focal point is typically around the center of the frame, therefore stabilization related distortions are less likely to be in this region. 
Second, since edges of frames are likely to be created through an inpainting process following stabilization, this ensures exclusion of those parts of frames from estimation.
And finally, by choosing the same location at each frame, corresponding PRNU extracts can be combined together to obtain a more reliable estimate.

\subsection{Warping Inversion}

Inversion of stabilization warps essentially corresponds to determining the correction applied to frames due to smoothing of estimated camera motion.
In the absence of an unstabilized, original video, this can only be realized by performing a blind search for corresponding transformation parameters at a given locality.
Obviously, the complexity of this task is determined by the nature of camera motion.
At the simplest, an affine motion model can be assumed. 
This will be effective when camera motion is only limited to translations and rotations. 
However, since an affine transformation preserves the parallelism of lines (but not their lengths and angles), it cannot correct perspective projections introduced by a moving camera. 
Hence, to take into account more complex stabilization transformations, projective transformations can be utilized.


A projective transformation can be represented by a $3\times 3$ matrix with 8 free parameters that specify the amount of rotation, scaling, translation, and projection applied to a point in two-dimensional space. 
In this sense, affine transformations form a subset of all such transformations without the two-parameter projection vector. 
Since a transformation is performed by multiplying the coordinate vector of a point with the transformation matrix; its inversion requires determining all these parameters. 
This problem is further exacerbated when transformations are applied in a spatially variant manner as different parts of the block might have undergone different transformations and when the block size is relatively small which yields to lower PCE values.






To determine the correct transformation applied to a block within a video frame, the block is inverse transformed repetitively and the transformation that yields the highest PCE between the inverse transformed block and the reference PRNU pattern is identified.  
In realizing this, rather than changing transformation parameters blindly which may lead to unlikely transformations and necessitate interpolation to take non-integer coordinates to integer values, we considered transformations that move corner vertices of the block within a search window.
In this regard, a large search window is preferable for more correct identification of the transform; however, search complexity grows polynomially with the size of the window. 

In determining a window size, we utilized several videos manually stabilized using the iMovie video editing program.
We observed that at 10\% stabilization setting, 
feature points move at most within a window of $15\times15$ pixels.
In fact, this observation aligns well with findings of earlier work in the field.
In \cite{LIU-SteadyFlow}, Liu {\em et al.} determined that considering dynamic scenes, points on tracked feature trajectories exhibit on average an unwanted motion of 2.36 pixels with a great majority of points moving less than 8.4 pixels overall.
Obviously, with increasing camera motion such deviations are likely to increase.
In \cite{grundmann-AutoDirected}, it is exhibited that low-frequency, up and down motions caused by walking may go up to 30 pixels. 
\textcolor{black}{Similarly, Iuliani {\em et al.}, by providing measurements obtained from several videos demonstrate that stabilization induced pixel movements can be within a range of $\pm24$ pixels (see variation in central cropping positions given in Table 3 of \cite{piva}).}

Therefore, a larger search window is expected to increase the chances of determining the correct stabilization transformation at the expense of considerably more computation.
In line with these observations, in our method, we assume that coordinates of each corner of a selected block may move independently within a window of $15\times15$ pixels, {\em i.e.,} spanning a range of $\pm 7$ pixels in both coordinates with respect to original position.
However, the block might have been subjected to a translation ({\em e.g.}, introduced by a cropping) not contained within the searched space of transformations. 
To be able to detect such translations as well, each inverse transformed block is also searched within a shift range of $\pm 50$ pixels in all directions in the reference pattern.


\textcolor{black}{
Instead of performing a pure random search over all transformation space, we adopted two approaches with different computational overheads
to accelerate the warping inversion step.
}

\subsubsection{Three-Level Hierarchical Grid Search (HGS)} 
\begin{figure}[!h]
	\centering
	\includegraphics[width=0.5\columnwidth]{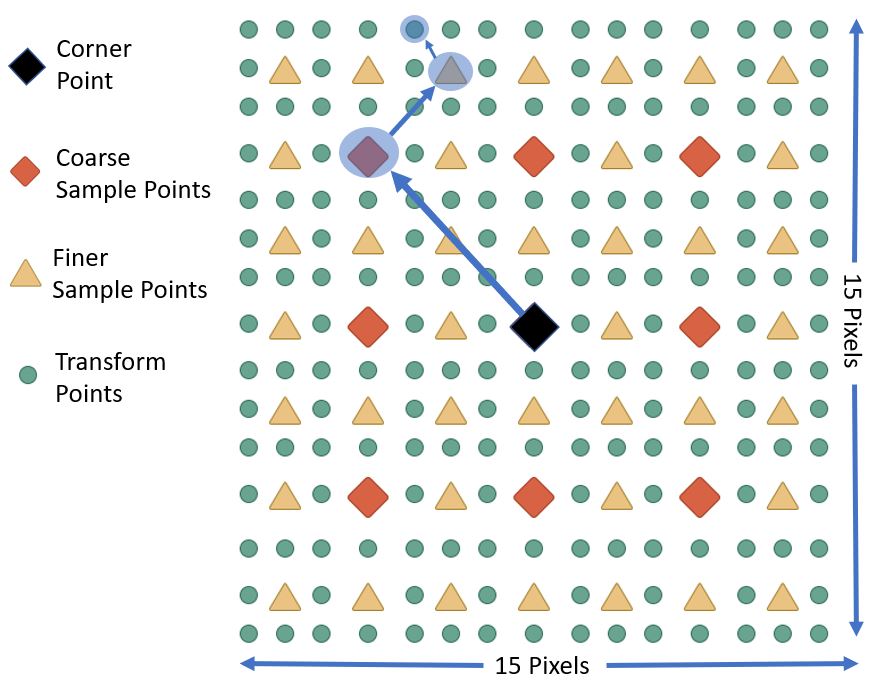}
	\caption{Three-level hierarchical grid partitioning of transform space. Coarse grid points that yield high PCE values are more finely partitioned for subsequent search.
	The arrows shows a sample trace of search steps to identify a likely transformation point for one of the corner vertices of a selected block.}
	\label{fig:search2}
\end{figure}

With this approach, the search space over rotation, scale, and projection is coarsely sampled. 
In the first level, each corner coordinate is moved by $\pm 4$ pixels (in all directions) over a coarse grid to identify five transformations (out of $9^4$ possibilities) that yield the highest PCE values. 
A higher-resolution search is then performed by the same process over neighboring areas of the identified transformations on a finer grid by changing the corner coordinates of transformed blocks $\pm 2$ and, again, retaining only the 5 transformations producing the five highest values. 
Finally, in the third level, coarse transformations determined in the previous level are further refined by considering all neighboring pixel coordinates (around a  $\pm 1$ range) to identify the most likely transformations needed for inverting the warping transformation due to stabilization. 
This overall reduces the number of transformations from $15^8$ to $11\times9^4$ possibilities, thereby yielding a significant reduction in complexity.
A pictorial depiction of the grid partitioning of the transform space is shown in Fig. \ref{fig:search2}.

\subsubsection{Constrained Three-Level Hierarchical Grid Search}
\label{sec:c3l}
\textcolor{black}{The search complexity can be further lowered by reducing the size of the $15\times 15$ pixels search window centered around each corner vertex. 
This can simply be realized by performing a two-level HGS which will essentially cover all translations within a $7\times 7$ search window, {\em i.e.,} movements up to $\pm 3$ pixels in both directions instead $\pm 7$ pixels as shown in Fig. \ref{fig:search2}. 
}
However, despite a significantly reduced transformation space, resulting complexity will still be in the same order as that of the three-level HGS, requiring $6\times9^4$ inverse transformations.

\textcolor{black}{
A computationally more efficient way of performing this search is to fix coordinates of one of the corner vertices in the three-level HGS by taking advantage of the subsequent search for global shifts.
(It must be noted that each inverse transformed block is further searched for possible translations of up to $\pm50$ pixels
in each direction.
Since this is realized through normalized cross-correlation and can be carried out very efficiently in the frequency domain, it does not pose an additional concern for computation overhead.) 
In essence, these shifts introduced to each inverse transformed block compensate for fixing one of the vertices of three-level HGS and  
ensure coverage of all transformations searched by the two-level HGS while providing almost an order of magnitude decrease in computation, from $11\times9^4$ to $11\times9^3$ as movements of only three vertices are considered.
}

\begin{figure}[!h]
	\centering
	\includegraphics[width=0.9\columnwidth]{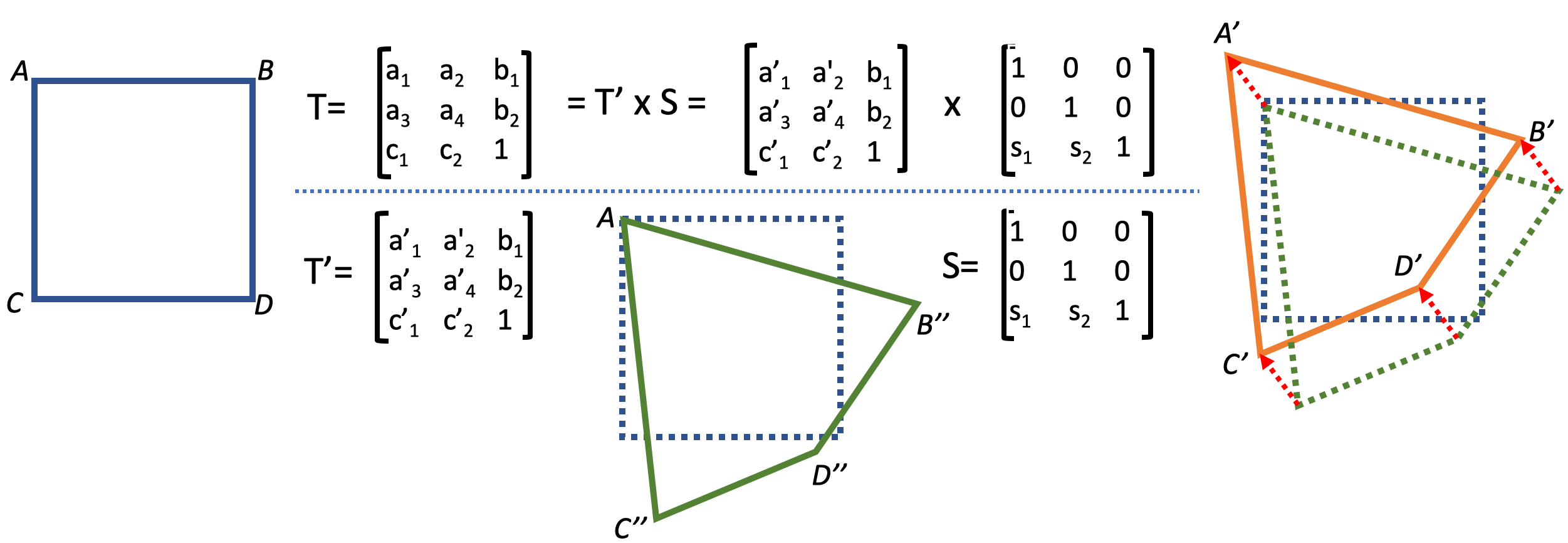}
	\caption{\textcolor{black}{
	Application of a stabilization transformation $T$ to a central PRNU block with corner vertices $(A,B,C,D)$ which transformed it into a quadrilateral with vertex coordinates at $(A', B', C', D')$. The transformed block can also be obtained by applying a transformation $T'$ that results with vertex coordinates at $(A, B'', C'', D'')$, which retains vertex A fixed, followed by a global shift in the direction of $A'$. } 
	}
	\label{fig:fixedPoint}
\end{figure}

\textcolor{black}{
This operation is depicted in Fig. \ref{fig:fixedPoint}.
Let $A=(X_a,Y_a)$, $B=(X_b,Y_b)$, $C=(X_c,Y_c)$ and $D=(X_d,Y_d)$ represent corner vertex coordinates of a PRNU block prior to application of an (exaggerated) stabilization transformation $T$ 
(blue rectangular block on the left). 
Given the warped block obtained after stabilization with corresponding vertex coordinates $A'=(X_a+x_a,Y_a+y_a)$, $B'=(X_b+x_b,Y_b+y_b)$, $C'=(X_c+x_c,Y_c+y_c)$ and $D'=(X_d+x_d,Y_d+y_d)$ (orange colored quadrilateral on the right), its shifted version in the direction of ($-x_a,-y_a$) will have coordinates
$A = (X_a,Y_a)$, $B''=(X_b+x_b-x_a,Y_b+y_b-y_a)$, $C''=(X_c+x_c-x_a,Y_c+y_c-y_a)$ and $D''=(X_d+x_d-x_a,Y_d+y_d-y_a)$ (green colored quadrilateral in the center).
Crucially, this shifted version can be viewed as the result of another warping, say, by some transformation $T'$ that preserved coordinates of the corner vertex $A$ of the original block.
That is to say, a generic transformation $T$ can be decomposed into a transformation $T'$ that retains one of the vertices fixed 
and a translation transformation $S$ that follows it, as shown in Fig. \ref{fig:fixedPoint}.
It must be noted, however, that the three-level HGS can only identify transformations that cause translations up to $\pm 7$ pixels in both coordinates, {\em i.e.,} $|x_{\varepsilon}|\leq7$ and $|y_{\varepsilon}|\leq7$ where $\varepsilon \in\{a,b,c,d\}$.
Therefore, the constrained search approach is limited to a subset of transformations covered by the three-level HGS that satisfy 
$|x_\varepsilon-x_a|\leq 7$ and $|y_\varepsilon-y_a|\leq 7$ for a fixed point $A$. 
Nevertheless, resulting transformations span the transformation space of the two-level HGS, where corner vertices $(A,B,C,D)$ can only move within a window of $7\times7$ pixels centered around them. \\
}



\subsection{Transform Validation}
Due to various factors, such as spatially variant nature of stabilization, coarse sampling of transformation space, small block size, and further weakening of PRNU patterns in video frames by compression and downsizing, 
very high PCE values may not be achieved even after correctly inverting the transformation. 
(It must be noted that the reliability of a PRNU pattern extracted from a video frame that underwent compression using the typical quantization parameter value of 20 is comparable to JPEG compression at quality factor of 65 \cite{H246H265}.)
Therefore, the warping inversion step might return an incorrect transformation due to a spurious match.
To assess the correctness of the identified inverse transformation, this step incorporates following additional measures. 
\begin{itemize}
    \item{$PCE_{vld}$}: A validation threshold is set to eliminate an unlikely transformation associated with a selected block. 
    When warping inversion yields a PCE value lower than $PCE_{vld}$, the identified transformation is assumed to be incorrect, and those blocks are excluded from PRNU estimation.
    To prevent false eliminations, this threshold must be set to a value well below the commonly accepted decision threshold used for photos, {\em i.e.,} PCE value of 60.   

    \item{$(n_{sub},PCE_{sub})$}: 
    It is possible that the inverse transformed PRNU block involves parts of frame content that underwent different warpings during stabilization.  
    To ensure that the identified transformation is largely prevalent over the block and is not due to some localized, content interference related phenomena, 
    we test whether the four non-overlapping $250\times250$ blocks comprising the central $500\times500$ block exhibit some coherence in the matching behavior. 
    \textcolor{black}{To realize this, the number of inverse transformed sub-blocks that yield a PCE value above a specified threshold ($PCE_{sub}$) are determined.  
    If this number is above a predetermined value ($n_{sub}$) the block is assumed to be correctly inverse transformed.} 

\end{itemize}
When the PCE values associated with a block and its sub-blocks are found to be below threshold values, {\em i.e.,} $PCE_{vld}$ and $PCE_{sub}$, a PRNU pattern could not be estimated from that particular frame.
In that case, warping inversion and transform validation steps are applied to remaining video frames. 

\subsection{Weighting \& PRNU Pattern Estimation}
In the last step, remaining PRNU patterns corresponding to central $500\times500$ block in each frame, following warping inversion and validation steps, are combined together to obtain a more reliable estimate of the sensor's PRNU. 
With videos, the reliability of the extracted PRNU pattern depends also on the level of compression applied during video coding.
Since encoder operates at a macroblock level by quantizing blocks at varying strengths, each macroblock's PRNU contribution can be weighted to take into account quantization related information loss to obtain a better estimate.
Hence, a pattern is estimated by simply taking an average of weighted PRNU patterns using the weighting scheme described in \cite{H246H265}.
This is essentially realized by creating a frame-level-mask with weighting coefficients determined based on the size, location, and quantization parameter of each macroblock. 
It must be noted that the mask corresponding to each frame is also transformed by the same transformation identified for a block.
The overall estimate obtained through PRNU weighting is then matched with the camera's reference PRNU to finally make an attribution decision. 

\section{Performance Evaluation}
\label{sec:experiments}

\textcolor{black}{To test the effectiveness of our method we used a standard dataset and two custom-built datasets}. 
\subsection{Datasets}

\subsubsection{VISION dataset} This dataset includes a collection of photos and videos captured by 35 different camera models \cite{shullani2017vision}. 
The videos in the dataset are divided into three sub-categories in terms of their content characteristics as having flat background, indoor, and outdoor scenes.
For each content sub-category three types of videos are acquired under increasing camera motion where the camera was still, moving, and manually panned and rotated.
(These different types of videos will be shortly referred to as still, move, or panrot videos.)
All videos are about 70 seconds long and initially acquired using the native camera application.
Out of the 35 cameras, only 16 of them performed stabilization with 13-32 videos available per camera.
This provided us with a total of 295 original videos captured by these cameras.
\textcolor{black}{Out of these videos, however, 38 of them had a low resolution (less than $1920 \times 1080$ pixels) and, therefore, were removed from further analysis.}  


\textcolor{black}{To measure the source camera verification accuracy, all of the $257$ full-HD, stabilized videos are used during tests. }
In all cases, the reference PRNU pattern for each camera is obtained using photos captured by the same camera.
\textcolor{black}{The amount of cropping and scaling applied to the full-frame sensor output to obtain video frames by in-camera downsizing are determined based on findings of \cite{luisapaper}, which reported both parameters for corresponding cameras in the VISION dataset.}
The tests are also repeated using non-matching reference PRNU patterns to measure the false-positive rate of the method.

Our method is devised to verify the source of stabilized videos where transformations are more complex than application of frame-level affine transformations.
\textcolor{black}{To identify those strongly stabilized videos, we set the thresholds for $stb_{chk}$ and $stb_{lite}$ tests to PCE values of 60 and 100, respectively.}
This resulted with elimination of the 105 out of the 257 tested videos by the $stb_{chk}$ test as they can already be successfully attributed to their sources. 
\textcolor{black}{
The remaining 152 videos are then subjected to $stb_{lite}$ test to identify weakly stabilized ones.
This is realized by aligning PRNU patterns of I frames with respect to the reference pattern obtained from photos through a search of transformation parameters considering
scaling factors around  the values determined in \cite{luisapaper} for all cameras, rotations up to $\pm1.5^{\circ}$ with $0.1^{\circ}$ increments, and all possible shift positions, as performed by \cite{piva} while setting aggregation threshold to a PCE value of 38.
Overall, 108 videos, out of 152, could be reliably attributed to their sources under a frame-level affine transformation model.
%
Hence, this left us with 44 strongly stabilized videos that cannot be attributed using earlier proposed approaches.} 
Examination of these videos revealed that, except for one, they belong to the panrot and move video categories where acquisition is performed under translational camera motion.
\textcolor{black}{An attempt to verify the source of these videos on a frame-by-frame basis using the basic method resulted with only two frames exceeding the PCE value of 60 (with values 78 and 491) while most frames yielded values less than 10.
This further showed that stabilization transformations were applied to almost all frames.}

\textcolor{black}{
\subsubsection{iPhone SE-XR Dataset}
This includes a collection of images and videos captured by two smartphone models, namely, iPhone Special Edition (SE) and iPhone XR.
These two models are selected due to two main reasons.
First, they are more recent smartphone models, and therefore, they plausibly deploy a state-of-the-art, in-camera stabilization solution. 
Second, both phone models feature only one main camera which prevents any potential post-processing related complications due to availability of multiple cameras (such as HDR/WDR processing, noise reduction, {\em etc.}).
This dataset includes media captured by 8 different phones (2 iPhone SE and 6 iPhone XR models) with a total of 263 photos and 41 videos (11-54 photos and 2-14 videos per camera).
These media are collected by searching for public Flickr profiles and using phones that we had access to.
Sources of Flickr photos are validated through pair-wise matching ($PCE>60$) and through metadata verification.
Another criterion we used for inclusion in the dataset was the availability of at least one still or low-motion video to ensure success of source verification and parameter estimation.} 

\textcolor{black}{
Following frame extraction with compensation for loop filtering and frame level PRNU pattern estimation, the two stabilization tests conducted.
Firstly, through the $stb_{chk}$ test it is determined that all videos are indeed stabilized. 
Then the reference PRNU patterns needed for source verification are obtained from photos; however unlike in the case of VISION dataset, the scaling factor used for 
downsizing photos to frames were not known a priori and had to be determined. 
\textcolor{black}{To this purpose}, the photo-based PRNU reference pattern and the PRNU pattern from the first frame of videos are matched considering 
rotations up to $\pm1.5^{\circ}$ with $0.1^{\circ}$ increments, all shift positions, and all scaling factors in the range of 0.3 to 0.9 with increments of 0.01.
The search for involved parameters revealed that iPhone XR and SE perform scaling with factors around 0.86 and 0.79, respectively. 
%
The $stb_{lite}$ test is then conducted as done for the VISION dataset using identified scaling factors for every video with 10 successive I frames following the first I frame. 
(For videos with less than 11 I frames, maximum number of available I frames are used.)
In cases where a match could not be attained, scaling factors obtained for other videos of the same camera model are also tried with the video in test. 
Overall, following the $stb_{lite}$ test, source of 19 videos could be verified and the remaining 22 videos are identified as strongly stabilized.}

%

\textcolor{black}{
\subsubsection{Adobe Premiere Pro Stabilized (APS) Video Dataset}
This dataset includes videos captured by 7 Android OS based smartphones through a custom camera application that turns off electronic stabilization and digital zoom and uses the camera's default compression setting \cite{ei2019}. 
It contains a total of 30 videos with 4-5 videos per camera.
Videos are shot indoors under a smooth panning motion with durations varying from 4-6 seconds. 
(The list of camera models and the number of available videos captured by each camera are given in Appendix \ref{app:a}.)
One of the videos from each camera is used to obtain a reference PRNU pattern while the remaining 23 videos are externally stabilized using Adobe Premiere Pro software suite. 
The choice of this software is due to its explicit indication of performing spatially variant stabilization \cite{adobe}.
Videos are stabilized at 10\% stabilization level while allowing frame-cropping but without post-stabilization scaling.
%
For these cameras, since the reference PRNU pattern can be obtained from a non-stabilized video, determining a scaling factor was not a concern.
Therefore, when performing the $stb_{lite}$ test only a slight scaling (with factors of $1\pm0.01$), rotations in the above specified range, and shifts within a range of $\pm$100 pixels are considered as parameters of affine transformations.  
It is verified through this test that out of the 23 videos 8 are weakly stabilized and 15 are strongly stabilized.
}


\textcolor{black}{
Table \ref{datasetsProperty} provides a brief description of the three video datasets in terms of their stabilization characteristics.
}

\begin{table}[!ht]
	\centering
	\caption{Dataset Features}
	\label{datasetsProperty}
	\begin{tabular}{c|c|c|c|c}
	&\# of   & \# of  & \# of weakly & \# of strongly  \\	
	Dataset&Cameras  & unstabilized  & stabilized & stabilized   \\
		&   & videos  & videos & videos  \\
		\hline
    VISION&14&105&108&44 \\ \hline
    iPhone SE-XR &8 & 0& 19 & 22 \\ \hline
    APS & 7 &0 & 8 &15 \\ \hline

	\end{tabular}
\end{table}

\subsection{\textcolor{black}{Performance of Three-Level HGS}}
\label{sec:sec-B}

Before performing attribution tests, a number of parameters related to our proposed approach must be determined.  
Most notably, this concerns the transform validation step which is necessary for eliminating transformations that are very likely to be incorrect.
To accept or reject an identified transformation associated with a block, resulting PCE value is first compared to the $PCE_{vld}$.
Then, a validation is performed at the sub-block level which includes an acceptance threshold for each sub-block and the minimum number of blocks that need to exceed this threshold, {\em i.e.,} $(n_{sub}, PCE_{sub})$.

\textcolor{black}{To determine these three parameters for each dataset, we utilized 5 frames from each of the strongly stabilized videos. 
Using the camera reference patterns and a set of arbitrary non-matching patterns, we performed transform inversion on central $500\times500$ blocks of all frames to identify the transformations that yield the best match. 
We then performed a sweep over the three parameters considering the whole range of values $0-40$ for $PCE_{vld}$, $0-4$ for $n_{sub}$, and $0-5$ for $PCE_{sub}$ 
that maximises correct identification rate while false identification among non-matching cases is set to zero.
Based on the observed accuracy values, best results are achieved when validation parameters are set to $PCE_{vld}=28$, $n_{sub}=2$, and $PCE_{sub}=2$ for all datasets except for the APS dataset for which $n_{sub}$ is found to be 1.
Overall, despite a slight variation, this finding shows that the determined parameter values generalize well over the three datasets.
To further verify that this is indeed the case, we repeated the same process by combining the 81 strongly stabilized videos from all datasets together and determining overall parameters jointly.
This yielded $PCE_{vld}=28$, $n_{sub}=2$, and $PCE_{sub}=2$, as expected. 
}

An important concern with attribution of stabilized videos is a misidentification of warping transformations due to typically low PCE values. 
To contain such occurrences, at each level of HGS, we keep track of top five transformations that yield highest PCE values rather than only retaining the best one. 
Hence, when evaluating the accuracy of the method, we consider the correct transformation to be among these transformations.
\textcolor{black}{In accordance with this, in the last step, five PRNU estimates are obtained rather than just one. 
The first estimate is obtained using the best of the top-five transformations for each frame, the second using the second best of top-five, and so on. 
Then, each resulting estimate is matched with the reference PRNU pattern.
Since identified transformations are expected to converge, an attribution decision is made only if three of the top-five identified transformations yield a PCE value above the designated threshold.
}

When verifying the source of a video, we utilized a number of frames from each given video.
Although attribution accuracy will improve with the number of frames, the computational complexity forbids using a large number of frames. 
\textcolor{black}{To this purpose}, we only utilized I frames which are used for prediction of other frames and, therefore, undergo a more favorable compression during coding. 
Further, leaving a temporal gap between frames makes warping inversion step less affected by video content. 
\textcolor{black}{
For our tests, we used 5 and 10 successive I frames from each video to evaluate the method's performance. 
In all videos, first I frames are excluded from attribution as some cameras either do not stabilize the first frame or perform a lighter stabilization than the subsequent frames.
When determining the correct attribution rate, each strongly stabilized video in the three datasets are matched with the reference pattern of corresponding camera using   parameters determined for each dataset ({\em i.e.}, $PCE_{vld}$, $n_{sub}$, $PCE_{sub}$).
To also measure the false attribution rate, these tests are then repeated on a collection of 189 videos, including all stabilized videos in the VISION dataset (152) and all strongly stabilized videos in the APS and iPhone SE-XR datasets (37), using the parameters associated with each dataset and by matching each video with a randomly selected, non-matching reference PRNU pattern.}


\begin{figure}[htbp]
  \begin{minipage}[b]{0.98\columnwidth}
    \centering
    \includegraphics[width=0.8\columnwidth]{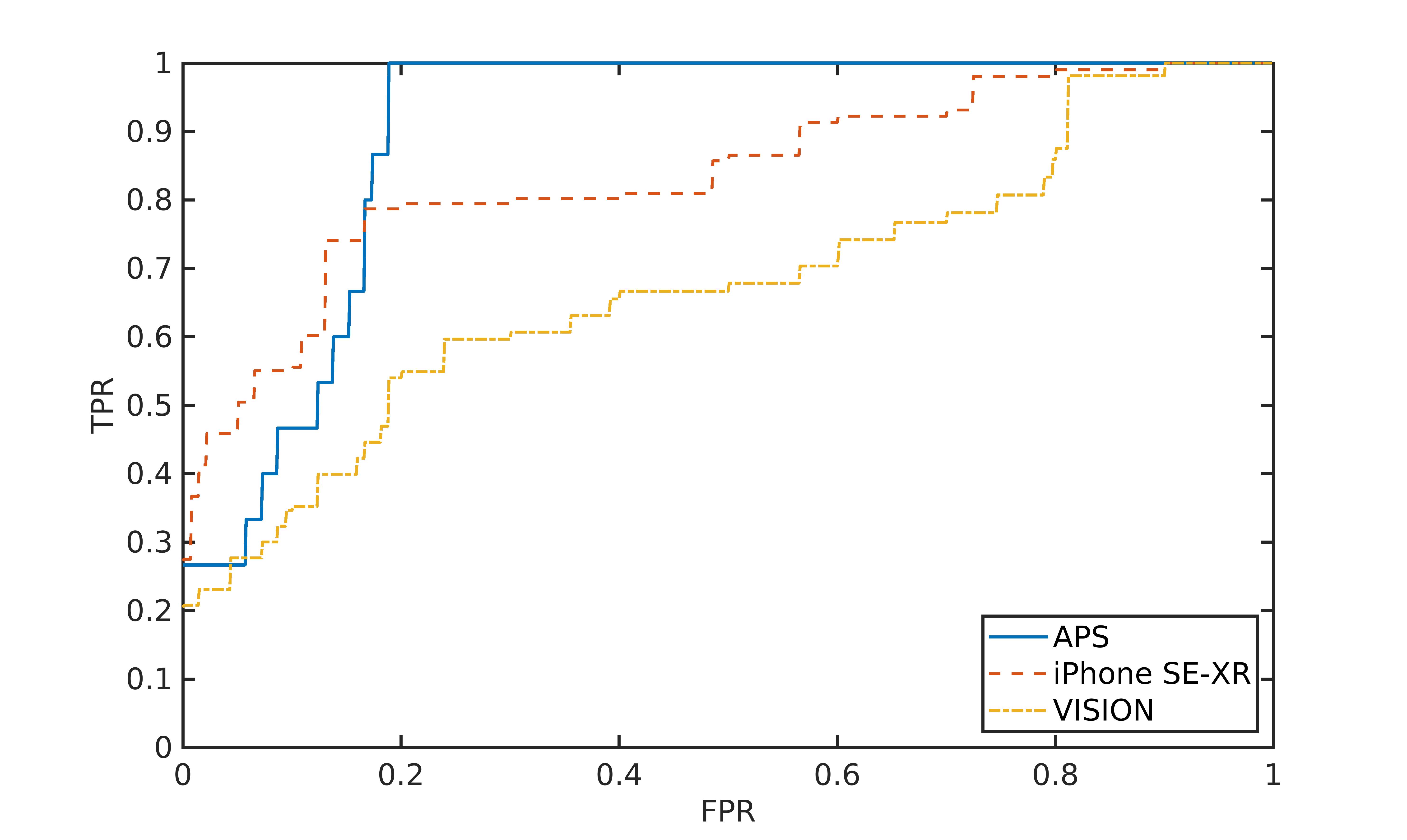}
    \centerline{(a)}
  \end{minipage}
  \begin{minipage}[b]{0.98\columnwidth}
    \centering
    \includegraphics[width=0.8\columnwidth]{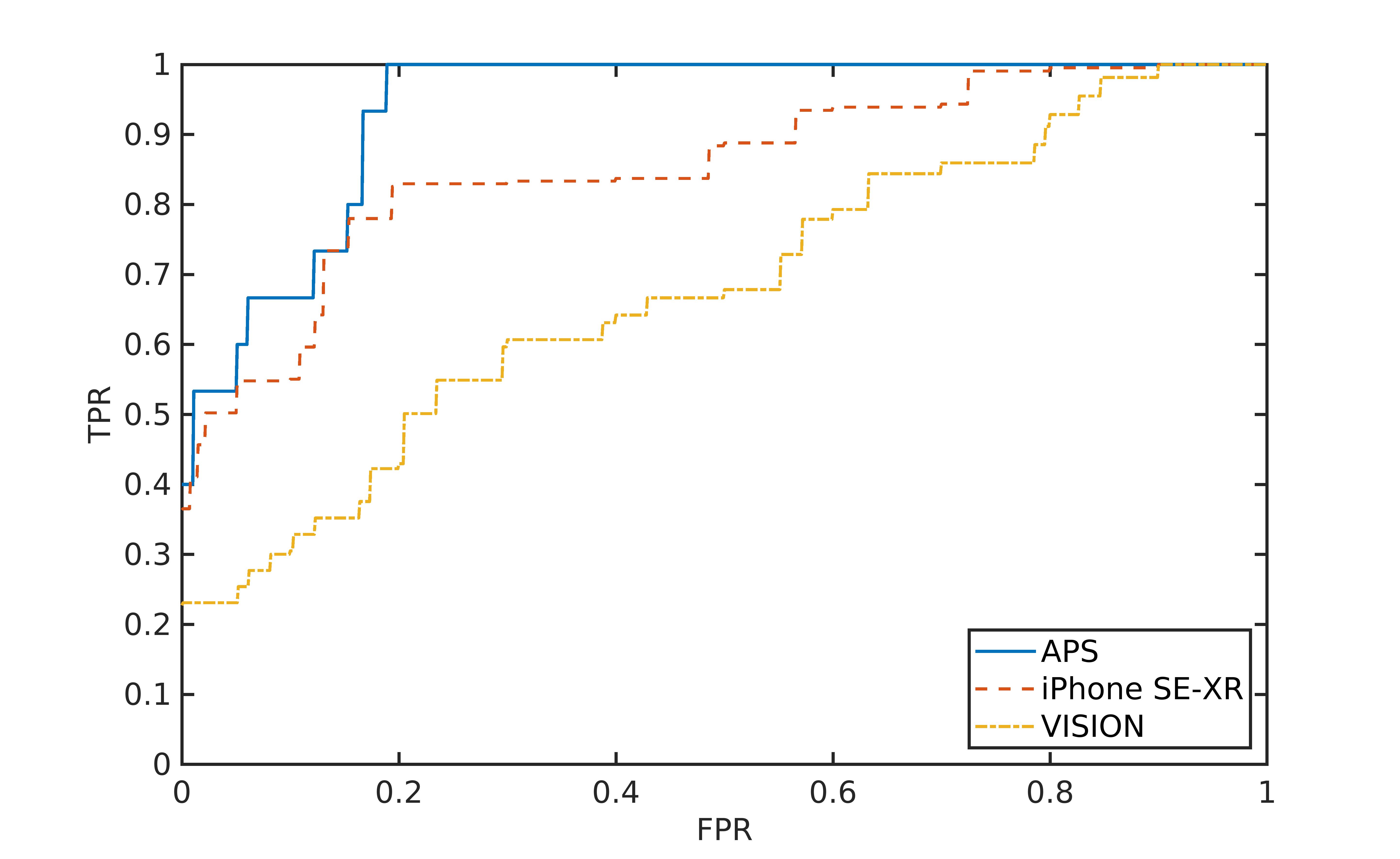}
    \centerline{(b)}
  \end{minipage}
	\caption{True and false positive attribution rates obtained for the 3 datasets when using (a) 5 and (b) 10 successive I frames from each video. Each pair of ROC curves is obtained using 15, 19, and 44 strongly stabilized videos in each dataset with matching sources and 189 stabilized videos with randomly selected, non-matching sources.}
	\label{fig:roc}
\end{figure}

Figure \ref{fig:roc} provides ROC curves obtained separately for each dataset to evaluate true and false positive rates (TPR and FPR) 
using different numbers of frames.
\textcolor{black}{
Each curve is generated by considering the 189 non-matching videos as negative samples and the videos belonging to a specific dataset ({\em i.e.}, APS, iPhone SE-XR, VISION) as positive samples. 
For evaluating TPR and FPR, we used the transform validation parameters ({\em i.e.}, $PCE_{vld}$, $n_{sub}$, $PCE_{sub}$) determined for each dataset.} 
In all cases, TPR and FPR values are obtained by comparing resulting PCE values (computed between the PRNU pattern obtained from each video and the corresponding reference PRNU pattern) to a varying threshold to make a decision.
During this computation, all videos that are eliminated at the transform validation step are assigned a PCE value of zero
to indicate that they do not match the reference pattern in question. 

It must be noted that some of the videos in the APS and iPhone SE-XR datasets do not have 10 I frames. 
For those videos results are obtained using all available I frames along with some randomly selected frames to reach 10 frames.
From the ROC curves it is determined that when only 5 frames are used for attributing videos in the APS, iPhone SE-XR, and VISION datasets, that cannot be attributed otherwise,
our method, respectively, achieves a TPR of 27\%, 27\%, and 20\% if FPR is set to 0\%.
Increasing the number of frames from 5 to 10 further improves the TPR in attribution of strongly stabilized videos to 40\%, 36\%, and 23\%, respectively, at 0\% FPR.

When evaluated together, these results show that, depending on whether 5 or 10 frames are used for attribution, proposed method is able to correctly attribute 19 to 24 of the 81 strongly stabilized videos in the three datasets, yielding a TPR of 23.4\% to 29.6\% at 0\% FPR.
Alternatively, when using jointly determined transform validation parameters, instead of parameters determined for each dataset, all but two of the those videos can be correctly attributed at 0\% FPR. 
Both of these videos are determined to be from the APS dataset. 
Overall, these results show that our method improves the state-of-the-art in attribution of strongly stabilized videos significantly.

We also \textcolor{black}{determine} the overall achievable attribution accuracy on these datasets.
Considering the 152 stabilized videos in the VISION dataset, incorporation of our end-to-end approach (see Fig. \ref{fig:process}) with existing methods focusing on attribution of weakly stabilized videos improves achievable attribution rate from 71\% to 77\%  when using 5 frames 
and to 77.6\% when using 10 frames, at 0\% FPR in both cases. 
(It must be noted here that earlier work, \cite{piva} and \cite{luisapaper} which mainly assume a frame-level affine transformatio model, reported results on VISION dataset considering videos captured by all cameras that support stabilization in their tests. 
Hence, the disparity between their results and the 71\% figure stems from the fact that our labeling of a video in the VISION dataset as stabilized or non-stabilized is mainly determined by $stb_{chk}$ test rather than inferring the label based on ability of the capturing camera to perform stabilization.)
Similarly for the iPhone SE-XR dataset overall attribution performance improved from 46\% to 61\% when using 5 frames and to 66\% when using 10 frames for attribution.
Finally, on the APS dataset overall accuracy increased from 35\% to 52\% with 5 frames and to 61\% with 10 frames.


 
\begin{figure}[!h]
	\centering
	\includegraphics[width=0.8\columnwidth]{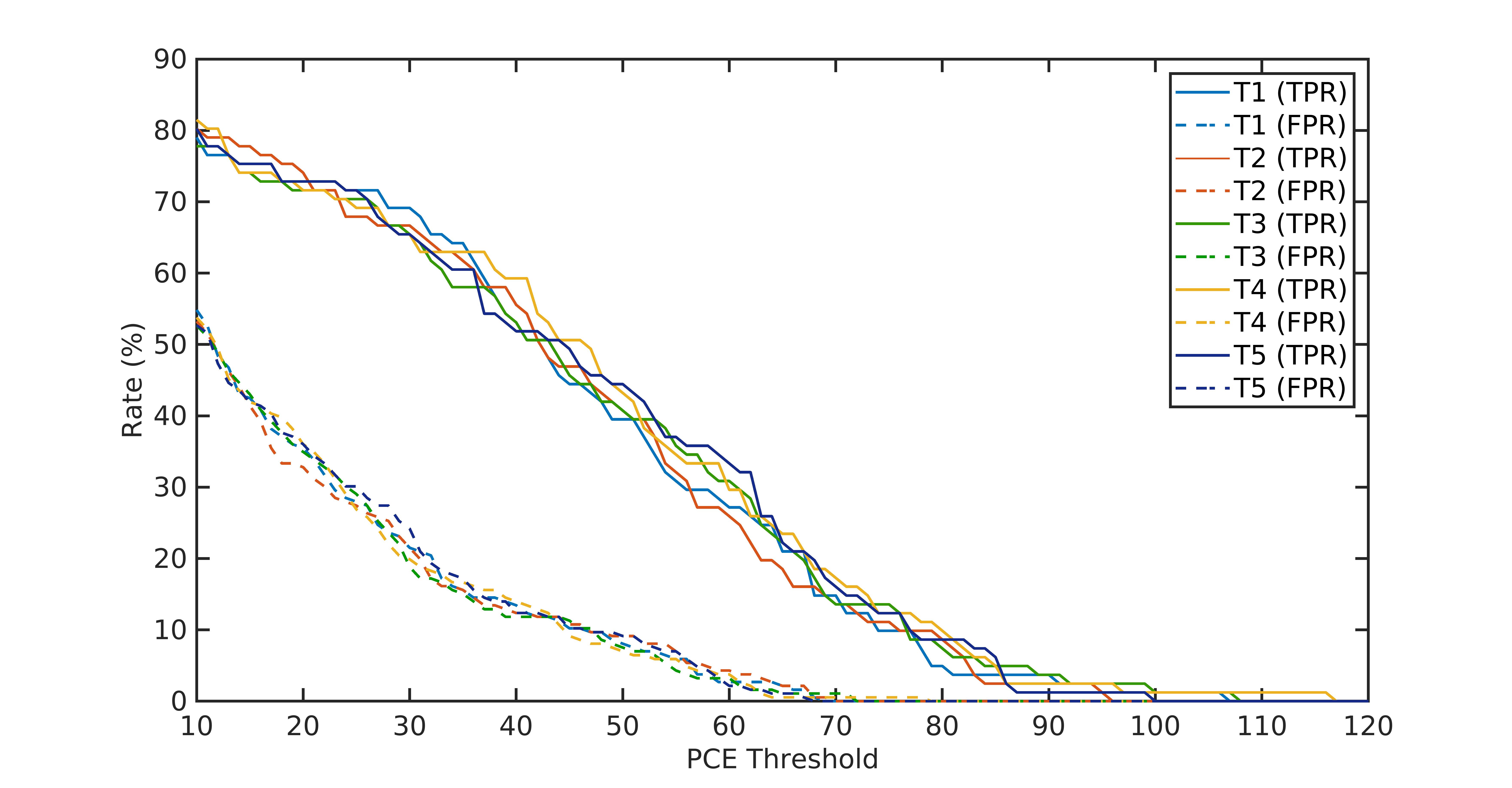}
	\caption{Change in attribution rates when top five identified transformations ($T_1,\ldots,T_5$) are used individually to make a decision. 
	Curves are obtained under the same test setting as in Fig. \ref{fig:roc} while utilizing only 5 frames from each video.}
	\label{fig:top5}
\end{figure}

We further \textcolor{black}{evaluate} how utilizing individual transformations among identified top-five transformations affects performance in making an attribution decision.
Figure \ref{fig:top5} shows TPR and FPR curves obtained for varying PCE threshold values using 5 I frames from 81 strongly stabilized videos and 189 non-matching videos where in each case only one of the top-five (inverse) transformations is used for attribution.   
It can be seen from these results that all transformations yields similar results. 
This result verifies that warping inversion does not identify arbitrary transformations and that it converges towards a very closely related set of transformations.
Therefore, the performance does not depend on which of the top-five transformations are used for making a decision.


An important component of our method is the transform validation step, introduced to eliminate incorrectly identified transformations.
Hence, this step is expected to be more effective in preventing false positive attributions.
However, since the PRNU block size of $500\times500$ is large enough to potentially include content that has undergone different transformations during stabilization, it will also eliminate some of the true positive matches.
To evaluate the impact of transform validation step on videos with matching and non-matching cases, we examined the number of frames eliminated by it.
Considering the earlier specified joint parameter values of $PCE_{vld}=28$, $n_{sub}=2$ and $PCE_{sub}=2$ 
and the 81 videos with matching sources, we determined that this step eliminated 1.69 frames on average when 5 frames are used for attribution.
Similarly, for the 189 videos with mis-matching sources, the average number of eliminated frames is found to be 3.5. 
In contrast, when 10 frames are used for attribution, on average 3.76 and 6.48 frames are eliminated for the source matching and non-matching cases, respectively.
This result reveals that transform validation step essentially eliminates many more frames of videos with mis-matching sources as compared to matching sources. 
As a consequence, this allows PRNU patterns extracted from various frames of a video to be combined together to obtain a more reliable PRNU estimate, which in turns yields a higher overall PCE value for correct attributions.

%
To complement this, we next tested the importance of the transformation validation step on the correct attribution performance.
For this, we first selectively removed the block ($PCE_{vld}=-\infty$) and sub-block ($n_{sub}=0,PCE_{sub}=-\infty$) level measures utilized by the validation step before eliminating it completely. 
In each case, the remaining measure is selected to achieve zero false attributions while maximizing correct attributions.
Table \ref{tab:TVS} provides the number of correctly attributed videos for the three cases. 
It can be seen from these results that just validating based on the $PCE_{vld}$ value is more effective than imposing ($n_{sub},PCE_{sub}$) values as it results with attribution 
of 12 videos as opposed to 6 videos attributed by using only the latter measure. 
If the transform validation step is removed completely, however, the number of attributed videos further drop to 5 videos from the initially attributed 19 videos. 
These results underline the key role of transform validation step in elimination of non-matching videos.


\begin{table}[!ht]
	\centering
\caption{Number of Correctly attributed videos With and Without Transform Validation Step Using 5 Frames}
 \label{tab:TVS}	
	\begin{tabular}{c|c|c|c}

\hline
Method & APS & iPhoneSE-XR & VISION\\ \hline
With Validation & 4 & 6 & 9 \\ \hline
$PCE_{vld}=-\infty$ & 1 & 2 & 3 \\ \hline
($n_{sub}=0,PCE_{sub}=-\infty$) & 2 & 3 & 7 \\ \hline
Without Validation & 1 & 1 & 3 \\ \hline

	\end{tabular}
\end{table}

To ensure that extracted PRNU patterns from the remaining frames contribute towards a more reliable PRNU estimate, we computed both the average PCE values corresponding to each block and the estimate reference pattern obtained by combining these individual PRNU patterns. 
For this, we examined all videos that have more than one frame that passed the transform validation step, regardless of whether the video is correctly stabilized or not.
When the sources matched ({\em i.e.,} for the 81 videos), we determined that the PCE values corresponding to inverse transformed blocks increases on average from 33 to 48 
when combined together.
In the case of successfully attributed videos, the average PCE value increases from 37.8 to 77.8 after combining PRNU patterns.
In the non-matching case, considering the 189 videos that yielded more than one transform validated frames, we determined that PCE decreases from 25 to 18. 
These results overall show that our method can obtain an estimate of the PRNU pattern under more complicated stabilization settings.

\subsection{\textcolor{black}{Constraining Complexity}}


\textcolor{black}{
The computational overhead of the three-level HGS is mainly determined by the number of transformations performed during the warping inversion step.}
In this regard, 
the three-level hierarchical grid partitioning of the transformation space while tracking multiple candidate points at each level 
yields a search complexity in the order of $9^4+5\times9^4+5\times9^4$ operations. 
Here, the term $9^4$ corresponds to the number of transformations evaluated when performing a search around a sample point at a given level (as exemplified in Fig. \ref{fig:search2}) and scalar terms determine the number of coarse or finer sample points this search is repeated at each level to identify most likely transformations.
Hence, improving the efficiency of the method requires alleviating the computational burden associated with these two terms.

The extent of the transformation space searched to identify the correct transformation is the main factor contributing to the complexity.
The three-level HGS essentially restricts the movement of each vertex to within a window of $15\times15$ pixels and performs a coarse sampling of the transformation space, reducing search space from $15^8$ to $11\times9^4$ transformations. 
\textcolor{black}{The constrained three-level HGS method will further decrease the complexity of the search.
However, as determined in Sec. \ref{sec:c3l}, this limits the search space to those stabilization transformations that induce relative movements (computed with respect to a fixed point in the center block) less than $\pm 7$ pixels along both coordinates.
Therefore, the feasibility of the constrained three-level HGS requires evaluating the trade-off between the computational gain and the verification accuracy due to reduced search space.}



\begin{figure}[htbp]
  \begin{minipage}[a]{0.49\columnwidth}
    \centering
    \includegraphics[width=0.95\columnwidth]{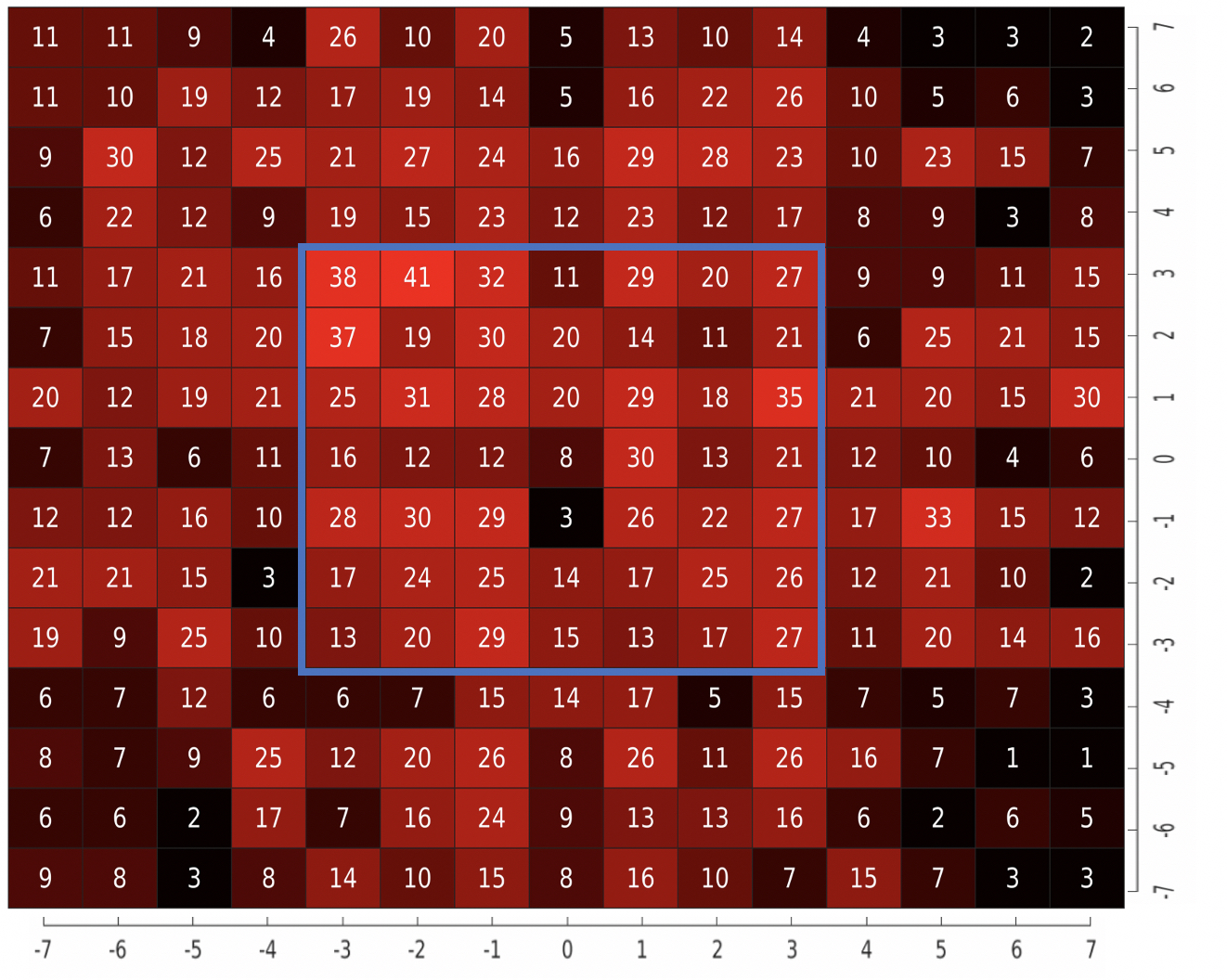}
    \centerline{(a)}
  \end{minipage}
  \begin{minipage}[a]{0.49\columnwidth}
    \centering
    \includegraphics[width=1\columnwidth]{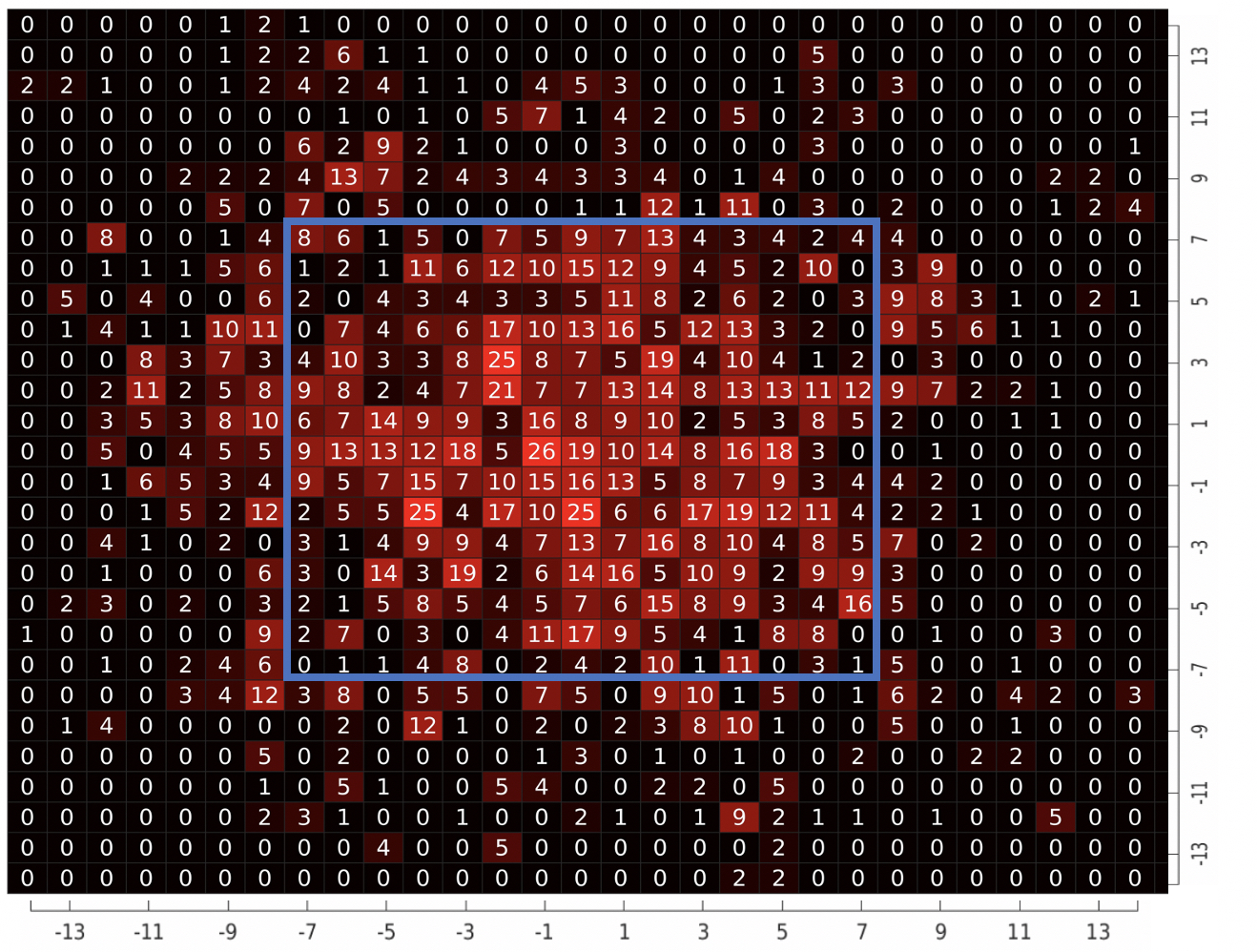}
    \centerline{(b)}
  \end{minipage}  

	\caption{\textcolor{black}{
	Heat maps showing observed stabilization induced movements for identified five most likely transformations associated with transform validated frames of correctly identified videos. 
	(a) \textcolor{black}{Movements of all corner vertices within a window of $15\times15$ pixels, centered around the original vertex position, corresponding to the search space of the three-level HGS. The blue rectangle shows movements within the $7\times7$ pixel window covered by the two-level HGS.}
	(b) Relative movements of corner vertices with respect to a fixed corner vertex of the same PRNU block. The blue rectangle shows the range of relative movements
	that are less than $\pm 7$ pixels on both axes.
	}
	}
	\label{fig:heatMap}
\end{figure}

To determine the potential impact on the performance, we examined the amount of unwanted motion compensated by image stabilization in all videos that were successfully matched to their sources.
\textcolor{black}{To this purpose}, movements in vertices of the central PRNU block of frames that pass the transform validation step for the five most likely transformations are determined. 
Figure \ref{fig:heatMap}-a visualizes the obtained distribution of values as a heat map while treating each vertex as an independent point.
\textcolor{black}{It is deduced from this figure that among all observed shifts that vary between $\pm 7$ pixels in both coordinates, only
33\% of the transformations yielded translations within a $\pm 3$ pixel range, {\em i.e.,} within the search space of the two-level HGS.
What is of more concern, however, is the relative movements of corner vertices of a PRNU block with respect to each other.} 
Figure \ref{fig:heatMap}-b displays the corresponding heat map for measured movements with respect to a fixed vertex. 
As expected, relative movements of vertices vary within a range of $\pm 14$ pixels; however, the distribution of observed movements are more clustered towards the center.
In fact, 73\% of movements are seen to be within a range of $\pm 7$ pixels which, when fully contains all movements, will allow the constrained search to attain the same level of matching performance as the three-level HGS.
Hence, this finding indicates that a favorable trade-off between computation gain and performance may be achieved.

The other factor that contributes to the computational complexity of the method relates to the additional search performed to identify the five most likely transformations instead of just one transformation. 
The three-level HGS essentially starts with a coarse sampling of transformation space and further refines it at each level, by performing a search of $9^4$ transformations in each level. 
However, to compensate for the error prone nature of the matching process, the second and third levels of search considers the five most likely candidate points from the preceding level to steer the search, effectively replicating the search effort at second and third levels by a factor of five. 
Hence, a computation gain may be achieved by tracking fewer number of candidate points.

\begin{figure}[htbp]
    \centering
    \includegraphics[width=0.4\columnwidth]{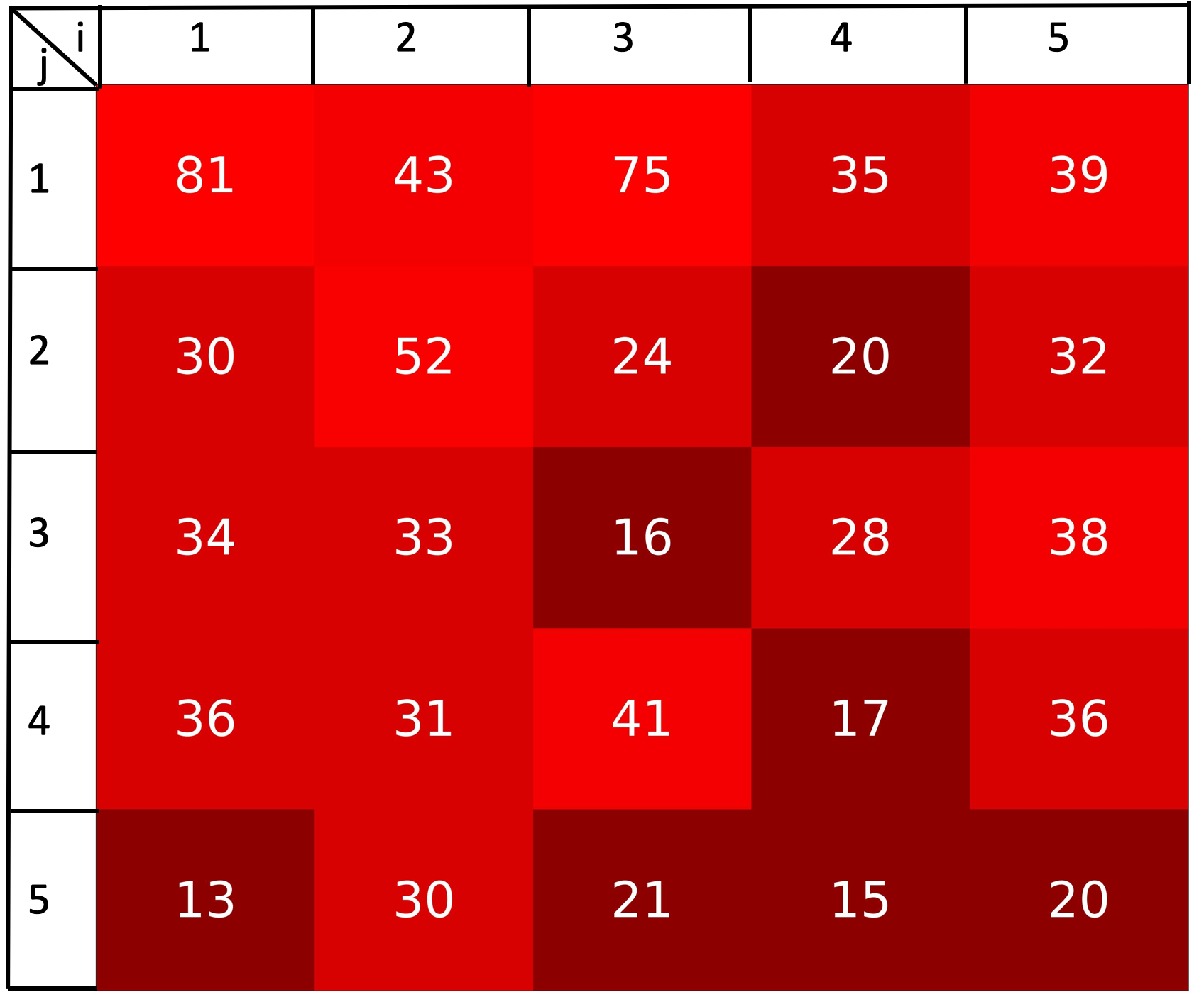}
	\caption{ 
	\textcolor{black}{
	Information on the origins of the five most likely transformations associated with frames of correctly matched videos obtained after the transform validation step. 
	Each element $(i,j)$ in the matrix denotes the number of transformations identified as a result of the search of transformations related to $i^{th}$ candidate points in the first level and $j^{th}$ candidate points in the second level.}
	}
	\label{fig:transformDist}
\end{figure}

The impact of such a change on the performance can be determined by examining the effectiveness of this tracking strategy.
\textcolor{black}{To this purpose}, we \textcolor{black}{investigate} how transformations identified at the third level of the search relates to the results of the search at first and second levels where multiple candidates are considered.  
Figure \ref{fig:transformDist} presents this information as a heat map where a value in location $(i,j)$ denotes the number of resulting transformations that are linked to $i^{th}$ candidate at the first level and $j^{th}$ candidate at the second level. 
(It must be noted that transformations identified for each individual frame are localized in the search space. 
This figure in contrast shows if there is a localization pattern that generalizes over all frames and videos.)
Accordingly, out of 840 transformations used for matching frames, only 81 of them would have been identified as top-five transformations if only the best candidates in the first and second levels were considered.
Overall, this observation indicates that considering fewer candidates at first level will cause a much higher decrease in the performance than decreasing the number of candidates in the second level.

\textcolor{black}{
To determine the interplay between computational gain and matching performance with respect to these factors, we repeated the same experiments on the three datasets. 
It must be noted that the constrained three-level HGS is expected to be less prone to false positives as it decreases the number of matches performed during verification from $9^4\times100^2$ to $9^3\times100^2$. 
Therefore, transform validation parameters must first be computed for the constrained search method.
To ensure the generalizability of results, parameters obtained for the VISION dataset are used across all datasets.
Following the same procedure described in Sec. \ref{sec:sec-B}, corresponding parameters are determined to be $PCE_{vld}=44$, $n_{sub}=2$, and $PCE_{sub}=2$. 
These values are very similar to those obtained for the three-level HGS with a notable increase in $PCE_{vld}$ from $28$ to $44$.
Examining this behavior, we determined that at this parameter setting 59\% of all non-matching videos are rejected directly as all of their frames are eliminated by the transform validation step. 
In comparison, for the three-level HGS, this ratio is found to be only 15\% for the corresponding parameters. 
This overall shows that decreasing the number of transformations, which in turn leads to a reduction in the number of false-positive matches, translates into an improved ability to eliminate non-matching videos. 
In the case of three-level HGS, we used the earlier determined parameters that yield the best results when performing the new experiments.
}

All tests are performed on a workstation with Intel Core i7-7700k processor and 32 GB memory running Ubuntu 18.04.1 LTS operating system. 
Table \ref{tab:results} provides a comparison of performance results, as well as computational overhead, for the three-level HGS and its constrained version considering the different number of candidate sample points at second and third levels. 
The second column of the table shows the number of times the search is repeated at each level, and the third column shows the overall number of inverse transformations performed by each method. 
The fourth column shows the corresponding time complexity of processing each frame.
The subsequent columns display the number of videos in each dataset whose source is correctly verified when 5 and 10 frames per video are used.

Accordingly, our end-to-end implementation of the verification method based on three-level HGS, which involves application of $72,171$ inverse transformations to a $500\times500$ sized PRNU noise block extracted from each frame (as detailed in Sec. \ref{sec:sec-B}), took on average 35 minutes per frame.
Restricting the number of candidate points in the second level and in both levels to only one reduced the number of correctly verified videos to 18 and 10 \textcolor{black}{(second and third rows of the three-level HGS)}, respectively, with a decrease in time complexity proportional to the reduction in the number of transformations performed. 
In both cases, however, the time complexity is found to be still relatively high, varying in the range 22-10 min/frame. 

\textcolor{black}{
Most critically, the constrained three-level HGS yielded a slightly lower source verification performance than the three-level HGS (19 videos as opposed to 24 \textcolor{black}{as presented in first rows of each method}) but resulted with almost an order of magnitude decrease in time complexity from 35 min/frame to 4 min/frame.
Among all datasets, the matching performance dropped notably only on the iPhone SE-XR dataset where the number of verified videos decreased from eight to three.
To explore this behavior, we examined the relative movements of the corner vertices associated with the eight videos matched by the three-level HGS, five of which were missed by the constrained search method.
We determined that for the latter five videos only about 65\% of vertex movements stayed within a range of $\pm 7$ pixels, whereas for the three videos that could be verified this ratio was 88\%.
This essentially demonstrates the advantage of the three-level HGS in covering a larger portion of the transformation space.
}


Finally, when both the number of candidate sample points at first and second levels and the search space are concurrently decreased \textcolor{black}{(second and third rows of the constrained three-level HGS)}, it is observed that the computation time reduces to a few minutes per frame; however, this gain realizes at the expense of a noticeable decrease in performance where only five videos could be matched to their sources.

\section{Discussion and Conclusions} \label{sec:Conc}

State-of-the-art stabilization methods pose a major challenge to source attribution of videos.
The difficulty mostly stems from the spatially variant nature of stabilization transformations which is further exacerbated by the adverse effects of in-camera processing steps, such as downsizing and video compression.  
Essentially, addressing this requires blindly inverting a geometric transformation while at the same time being restricted to operate on smaller blocks with significantly weakened PRNU patterns.
Our findings in this work \textcolor{black}{showed} that under strong stabilization, reliable estimation of a PRNU pattern from a video is not viable. 
Instead, the problem can be addressed in a source verification setting, where the match of a video with a known camera is in question.

\renewcommand{\tabcolsep}{3pt}
\begin{table}[!ht]
	\centering
	\caption{Comparison of Matching Accuracy and Computation Speed for Various Warping Inversion Approaches}
	\label{tab:results}
	\begin{tabular}{c|c|c|c|c|c|c|c|c|c}
\hline
 & Number of & \multicolumn{2}{c|}{Computation}  &  \multicolumn{2}{c|}{APS} & \multicolumn{2}{c|}{iPhone} & \multicolumn{2}{c}{VISION} \\
 Method & Searches  & Load &  Speed &   \multicolumn{2}{c|}{ } & \multicolumn{2}{c|}{SE-XR} & \multicolumn{2}{c}{ } \\ \cline{5-10}
  & {\tiny{\em (levels 1-3)}} &{\tiny{\em (\# of transformations)}}&{\tiny {\em (min/frame)}} &  5F & 10F & 5F & 10F &  5F & 10F\\ \hline
Three-Level & 1+5+5 & $11 \times 9^4$ & 35 & 4 & 6 & 6 & 8 & 9 & 10 \\ \cline{2-10}
HGS &1+5+1 & $7 \times 9^4$ & 22 & 4 & 4 & 5 & 8 & 6 & 6 \\ \cline{2-10}
  &1+1+1 & $3 \times 9^4$ & 10 & 2 & 2 & 2 & 3 & 2 & 5 \\ \hline
Constrained& 1+5+5 & $11 \times 9^3$ & 4 & 4 & 6 & 2 & 3 & 8 & 10 \\ \cline{2-10}
 Three-Level&1+5+1 & $7 \times 9^3$ & 2.5 & 1 & 2 & 0 & 0 & 3 & 3 \\ \cline{2-10}
 HGS&1+1+1 & $3 \times 9^3$ & 1 & 1 & 2 & 0 & 0 & 3 & 3 \\ \hline
	\end{tabular}
\end{table}

There is little information on details of stabilization methods deployed by cameras; therefore, it is not trivial to design methods that can provide optimal solutions with low computational cost.
Although estimation of PRNU patterns through inversion of frame-level affine transformations constitute the first step of a solution for attribution, 
as demonstrated by our results, this approach is not adequately effective on all stabilized videos in the VISION and iPhone SE-XR datasets.
To address this gap, our approach in this work builds on existing work in the field and expands capabilities to address more complicated forms of stabilization 
by assuming more degrees of freedom in the involved transformations and taking into account spatially variant nature of modern stabilization approaches.
Results obtained on strongly stabilized videos in the VISION, iPhone SE-XR, and APS datasets show that our three-level HGS method, respectively, achieves a source verification accuracy of 20-23\%, 27-36\%, and 27-40\% with no false-positive matches when using 5-10 frames for attribution.
For the constrained three-level HGS method, which provides about an order of magnitude reduction in computational time (35 min/frame versus 4 min/frame), 
the accuracy remains almost the same for the APS and the VISION datasets and decreases to 9-14\% for the iPhone SE-XR dataset.

The novelty of our method mainly stems from tackling the spatially variant nature of stabilization methods by searching a large range of projective transformations at sub-frame level and due to its ability to eliminate incorrectly identified transformations. 
Although our method improves existing capabilities considerably, reliable attribution of strongly stabilized videos requires further exploration.
One potential improvement area concerns obtaining further specifics about the stabilization methods deployed by cameras in smartphone type computing devices.  
Such an information can be translated into devising more effective warping inversion methods. 
\textcolor{black}{In the absence of such information, an immediate research direction is to develop improved strategies to search the stabilization transformation space.}
Another advancement that will help achieve better results is about reliable estimation of PRNU patterns.
Since with video frames smaller block sizes yield very weak PRNU patterns, overcoming this obstacle will have a direct impact on the success of warping inversion step.
Recently proposed deep learning based approaches \cite{cnnBase} can be considered a step in this direction.

Finally, we note that since transform inversion is done in a blind manner, incorrect identification of transformations is more likely to occur with increasing search space.
\textcolor{black}
{It must be noted that for each matching attempt, even the constrained search method performs millions of matches ($9^3\times100^2$) before identifying the correct stabilization transformation associated with a block and 
the potential translation applied to it.
}
Therefore, transform validation step is of vital importance to our method. 
Initially, when determining transformation parameters, we also considered imposing a continuity constraint between transformations applied to successive frames as the camera motion cannot change abruptly from one frame to another.
Our analysis, however, revealed that even videos with non-matching sources exhibit this characteristic.
That is, PRNU patterns extracted from two successive frames under very similar transformations may also yield similar PCE values with a non-matching reference PRNU pattern.
We conjecture that this behavior is mainly due to the content interference in the estimated PRNU pattern of successive frames.
Hence, it is necessary to sample frames from different parts of a video to suppress content dependency effects.

\section{Acknowledgement}
This work is supported by the Scientific and Technological Research Council of Turkey (TUBITAK) grant 116E273. We also thank E. S. Tandogan for his help in conducting some of the experiments.

%

%
\appendix
\section{List of camera models }
\label{app:a}

\textcolor{black}{
The camera models used for capturing videos in the APS dataset under controlled settings \cite{ei2019} ({\em i.e.} with stabilization and electronic zoom turned off and at default video compression settings) are listed below.
The numbers in parentheses denote the number of videos available from each camera.
}
\begin{table}[!ht]
	\centering
	\label{CameraModel}
	\begin{tabular}{c|c|c|c}
	
\hline
Samsung Edge 6 (5)& Redmi 5 Plus (5) & LG G4 (4) & Galaxy S4 (4)\\ \hline
Huawei P20 Lite (4)&LG G Flex2 (4) &LG G3 (4)&  \\ \hline

	\end{tabular}
\end{table}

\bibliographystyle{IEEEtran}
\bibliography{bibfile}

\begin{thebibliography}{10}
\providecommand{\url}[1]{#1}
\csname url@samestyle\endcsname
\providecommand{\newblock}{\relax}
\providecommand{\bibinfo}[2]{#2}
\providecommand{\BIBentrySTDinterwordspacing}{\spaceskip=0pt\relax}
\providecommand{\BIBentryALTinterwordstretchfactor}{4}
\providecommand{\BIBentryALTinterwordspacing}{\spaceskip=\fontdimen2\font plus
\BIBentryALTinterwordstretchfactor\fontdimen3\font minus
  \fontdimen4\font\relax}
\providecommand{\BIBforeignlanguage}[2]{{%
\expandafter\ifx\csname l@#1\endcsname\relax
\typeout{** WARNING: IEEEtran.bst: No hyphenation pattern has been}%
\typeout{** loaded for the language `#1'. Using the pattern for}%
\typeout{** the default language instead.}%
\else
\language=\csname l@#1\endcsname
\fi
#2}}
\providecommand{\BIBdecl}{\relax}
\BIBdecl

\bibitem{chen}
M.~Chen, J.~Fridrich, M.~Goljan, and J.~Luk{\'a}s, ``Determining image origin
  and integrity using sensor noise,'' \emph{IEEE Transactions on information
  forensics and security}, vol.~3, no.~1, pp. 74--90, 2008.

\bibitem{PCE}
\BIBentryALTinterwordspacing
B.~V. K.~V. Kumar and L.~Hassebrook, ``Performance measures for correlation
  filters,'' \emph{Appl. Opt.}, vol.~29, no.~20, pp. 2997--3006, Jul 1990.
  [Online]. Available: \url{http://ao.osa.org/abstract.cfm?URI=ao-29-20-2997}
\BIBentrySTDinterwordspacing

\bibitem{corcoran2016consumer}
P.~Corcoran, P.~Bigioi, J.~Chen, W.~Cranton, and M.~Fihn, ``Consumer imaging
  i--processing pipeline, focus and exposure,'' pp. 1--25, 2016.

\bibitem{zhang2018pixel}
J.~Zhang, J.~Jia, A.~Sheng, and K.~Hirakawa, ``Pixel binning for high dynamic
  range color image sensor using square sampling lattice,'' \emph{IEEE
  Transactions on Image Processing}, vol.~27, no.~5, pp. 2229--2241, 2018.

\bibitem{guo2018efficient}
J.~Guo, H.~Gu, and M.~Potkonjak, ``Efficient image sensor subsampling for
  dnn-based image classification,'' in \emph{Proceedings of the International
  Symposium on Low Power Electronics and Design}, 2018, pp. 1--6.

\bibitem{PRNUvideoJessica}
M.~Chen, J.~Fridrich, M.~Goljan, and J.~Luk{\'a}{\v{s}}, ``Source digital
  camcorder identification using sensor photo response non-uniformity,'' in
  \emph{Security, Steganography, and Watermarking of Multimedia Contents IX},
  vol. 6505.\hskip 1em plus 0.5em minus 0.4em\relax International Society for
  Optics and Photonics, 2007, p. 65051G.

\bibitem{MACEfilter}
D.-K. Hyun, C.-H. Choi, and H.-K. Lee, ``Camcorder identification for heavily
  compressed low resolution videos,'' in \emph{Computer Science and
  Convergence}.\hskip 1em plus 0.5em minus 0.4em\relax Springer, 2012, pp.
  695--701.

\bibitem{IPBdiff}
W.-H. Chuang, H.~Su, and M.~Wu, ``Exploring compression effects for improved
  source camera identification using strongly compressed video,'' in
  \emph{Image Processing (ICIP), 2011 18th IEEE International Conference
  on}.\hskip 1em plus 0.5em minus 0.4em\relax IEEE, 2011, pp. 1953--1956.

\bibitem{eusipco18}
E.~Altinisik, K.~Tasdemir, and H.~T. Sencar, ``Extracting prnu noise from h.264
  coded videos,'' in \emph{2018 26th European Signal Processing Conference
  (EUSIPCO)}.\hskip 1em plus 0.5em minus 0.4em\relax IEEE, 2018, pp.
  1367--1371.

\bibitem{H246H265}
E.~Alt{\i}n{\i}{\c{s}}{\i}k, K.~Ta{\c{s}}demir, and H.~T. Sencar, ``Mitigation
  of h.264 and h.265 video compression for reliable prnu estimation,''
  \emph{IEEE Transactions on information forensics and security}, 2019.

\bibitem{ei2019}
E.~S. Tandogan, E.~Alt{\i}n{\i}s{\i}k, S.~Sarimurat, and H.~T. Sencar,
  ``Tackling in-camera downsizing for reliable camera id verification,'' in
  \emph{Electronic Imaging, Media Watermarking, Security, and Forensics}.\hskip
  1em plus 0.5em minus 0.4em\relax Society for Imaging Science and Technology,
  2019.

\bibitem{piva}
M.~Iuliani, M.~Fontani, D.~Shullani, and A.~Piva, ``Hybrid reference-based
  video source identification,'' \emph{Sensors}, vol.~19, no.~3, p. 649, 2019.

\bibitem{taspinar2019source}
S.~Taspinar, M.~Mohanty, and N.~Memon, ``Source camera attribution of
  multi-format devices,'' \emph{arXiv preprint arXiv:1904.01533}, 2019.

\bibitem{taspinar2016}
------, ``Source camera attribution using stabilized video,'' in \emph{2016
  IEEE International Workshop on Information Forensics and Security
  (WIFS)}.\hskip 1em plus 0.5em minus 0.4em\relax IEEE, 2016, pp. 1--6.

\bibitem{luisapaper}
S.~Mandelli, P.~Bestagini, L.~Verdoliva, and S.~Tubaro, ``Facing device
  attribution problem for stabilized video sequences,'' \emph{IEEE Transactions
  on Information Forensics and Security}, 2019.

\bibitem{ref-Content2009}
F.~Liu, M.~Gleicher, H.~Jin, and A.~Agarwala, ``Content-preserving warps for 3d
  video stabilization,'' in \emph{ACM Transactions on Graphics (TOG)}, vol.~28,
  no.~3.\hskip 1em plus 0.5em minus 0.4em\relax ACM, 2009, p.~44.

\bibitem{dataset}
D.~Shullani, M.~Fontani, M.~Iuliani, O.~Al~Shaya, and A.~Piva, ``Vision: a
  video and image dataset for source identification,'' \emph{EURASIP Journal on
  Information Security}, vol. 2017, no.~1, p.~15, 2017.

\bibitem{apple}
\BIBentryALTinterwordspacing
``Developer.apple.com,'' 2020, January. [Online]. Available:
  \url{https://developer.apple.com/library/archive/documentation/DeviceInformation/Reference/iOSDeviceCompatibility/Cameras/Cameras.html#}
\BIBentrySTDinterwordspacing

\bibitem{elTitremesi}
B.~Golik and D.~Wueller, ``Measurement method for image stabilizing systems,''
  in \emph{Digital Photography III}, vol. 6502.\hskip 1em plus 0.5em minus
  0.4em\relax International Society for Optics and Photonics, 2007, p. 65020O.

\bibitem{Xu-FastFeature}
J.~Xu, H.-w. Chang, S.~Yang, and M.~Wang, ``Fast feature-based video
  stabilization without accumulative global motion estimation,'' \emph{IEEE
  Transactions on Consumer Electronics}, vol.~58, no.~3, pp. 993--999, 2012.

\bibitem{grundmann-AutoDirected}
M.~Grundmann, V.~Kwatra, and I.~Essa, ``Auto-directed video stabilization with
  robust l1 optimal camera paths,'' in \emph{Proceedings of CVPR 2011}.\hskip
  1em plus 0.5em minus 0.4em\relax IEEE, 2011, pp. 225--232.

\bibitem{patent_stab_gyro}
D.~J. Thivent, G.~E. Williams, J.~Zhou, R.~L. Baer, R.~Toft, and S.~X.
  Beysserie, ``Combined optical and electronic image stabilization,'' May~22
  2018, uS Patent 9,979,889.

\bibitem{LIU-SteadyFlow}
S.~Liu, L.~Yuan, P.~Tan, and J.~Sun, ``Steadyflow: Spatially smooth optical
  flow for video stabilization,'' in \emph{Proceedings of the IEEE Conference
  on Computer Vision and Pattern Recognition}, 2014, pp. 4209--4216.

\bibitem{360Kopf}
J.~Kopf, ``360 video stabilization,'' \emph{ACM Transactions on Graphics
  (TOG)}, vol.~35, no.~6, p. 195, 2016.

\bibitem{Wang-HiQuRealTime}
Z.~Wang, L.~Zhang, and H.~Huang, ``High-quality real-time video stabilization
  using trajectory smoothing and mesh-based warping,'' \emph{IEEE Access},
  vol.~6, pp. 25\,157--25\,166, 2018.

\bibitem{JessicaCrop}
M.~Goljan and J.~Fridrich, ``Camera identification from cropped and scaled
  images,'' in \emph{Security, Forensics, Steganography, and Watermarking of
  Multimedia Contents X}, vol. 6819.\hskip 1em plus 0.5em minus 0.4em\relax
  International Society for Optics and Photonics, 2008, p. 68190E.

\bibitem{karakucuk2015}
A.~Karak{\"u}c{\"u}k, A.~E. Dirik, H.~T. Sencar, and N.~D. Memon, ``Recent
  advances in counter prnu based source attribution and beyond,'' in
  \emph{Media Watermarking, Security, and Forensics 2015}, vol. 9409.\hskip 1em
  plus 0.5em minus 0.4em\relax International Society for Optics and Photonics,
  2015, p. 94090N.

\bibitem{down2}
L.~Bondi, P.~Bestagini, F.~Perez-Gonzalez, and S.~Tubaro, ``Improving prnu
  compression through preprocessing, quantization, and coding,'' \emph{IEEE
  Transactions on Information Forensics and Security}, vol.~14, no.~3, pp.
  608--620, 2018.

\bibitem{Goljan-LP}
M.~Goljan, ``Blind detection of image rotation and angle estimation,''
  \emph{Electronic Imaging}, vol. 2018, no.~7, pp. 1--10, 2018.

\bibitem{ilkIFrameNonStabil}
M.~Grundmann, V.~Kwatra, and I.~Essa, ``Cascaded camera motion estimation,
  rolling shutter detection, and camera shake detection for video
  stabilization,'' Feb.~6 2018, uS Patent 9,888,180.

\bibitem{shullani2017vision}
D.~Shullani, M.~Fontani, M.~Iuliani, O.~Al~Shaya, and A.~Piva, ``Vision: a
  video and image dataset for source identification,'' \emph{EURASIP Journal on
  Information Security}, vol. 2017, no.~1, p.~15, 2017.

\bibitem{adobe}
\BIBentryALTinterwordspacing
``Adobe premiere,'' 2020, January. [Online]. Available:
  \url{https://www.adobe.com/products/premiere.html}
\BIBentrySTDinterwordspacing

\bibitem{cnnBase}
M.~Kirchner and C.~Johnson, ``Spn-cnn: Boosting sensor-based source camera
  attribution with deep learning,'' in \emph{2019 IEEE International Workshop
  on Information Forensics and Security (WIFS)}.\hskip 1em plus 0.5em minus
  0.4em\relax IEEE, 2019.

\end{thebibliography}

\onecolumn
\label{sec:appendix}
\end{document}